%% file: Submitted_arxiv_20.9.2015.tex
\documentclass{ws-rv975x65}
\usepackage{subfigure}     
\usepackage{ws-rv-thm}     
\usepackage{ws-rv-van}     
\makeindex

\include{commands}

\begin{document}

\chapter[Precision Experiments at LEP]{Precision Experiments at LEP}\label{ra_ch9}

\author[W. de Boer]{W. de Boer}

\address{Karlsruhe Institute of Technology,\\
Institut f\"ur Experimentelle Kernphysik, Gaedestr. 1,  76131 Karlsruhe, Germany \\
wim.de.boer@kit.edu}

\begin{abstract}
The Large Electron Positron Collider (LEP)  established  the Standard Model (SM) of particle physics with unprecedented precision, including all its radiative corrections. These led to predictions for  the masses of the top quark and Higgs boson, which were beautifully confirmed later on.  After these precision measurements the Nobel Prize in Physics   was awarded  in 1999 jointly to 't Hooft and Veltman  
``for elucidating the quantum structure of electroweak interactions in physics''. 

Another hallmark of the LEP results were the precise  measurements of the gauge coupling constants, which  excluded unification of  the forces within the SM, but allowed unification within the supersymmetric extension of the SM. This increased the interest in Supersymmetry (SUSY) and Grand Unified Theories, especially since the SM has no candidate for the elusive dark matter, while Supersymmetry provides an excellent candidate for dark matter. In addition,  Supersymmetry  removes the quadratic divergencies of the SM and {\it predicts} the Higgs mechanism from radiative electroweak symmetry breaking with a SM-like Higgs boson having a mass below 130 GeV in agreement with the Higgs boson discovery at the LHC.   However, the predicted SUSY particles have not  been found either because they are too heavy for the present LHC energy and luminosity or Nature has found alternative ways to circumvent the shortcomings of the SM.
\end{abstract}

\clearpage
\body
\section{Introduction}\label{s1}
The Standard Model is a relativistic quantum field theory describing the strong and electroweak interactions of  quarks and leptons, which up to now are considered to be elementary particles. 
The complexity and non-triviality of the Standard Model (SM) of particle physics  is  lucidly described in the  36 Nobel  Lectures unra\-veling  the stepwise discovery of the SM in a personal way \cite{nobelall}.
The first example of a relativistic quantum field theory was quantum electrodynamics, which describes the electromagnetic interactions by the exchange of a massless photon.   The short range of the  weak interactions implies that they are mediated by   massive gauge bosons, the W- and Z bosons, which were discovered at the SPS, as described elsewhere in this volume. 

Relativistic quantum field theories based on local gauge symmetries had two basic problems: i) explicit gauge boson mass terms are not allowed in the SM, since they break the symmetry and ii) the high energy behaviour leads to infinities in the cross sections, masses and couplings. The first problem was solved  in 1964 by Higgs and others \cite{Higgs:1964ia,Higgs:1964pj,Englert:1964et,Guralnik:1964eu}, who proposed that gauge boson masses are generated by interactions with an omnipresent scalar (Higgs) field in the vacuum, so no explicit mass terms are needed in the Lagrangian for these dynamically generated masses. The quantum of the Higgs field, the Higgs boson, was discovered at the LHC in 2012, as described elsewhere in this volume.  After this discovery Englert and Higgs were awarded the Nobel prize in 2013. The second problem was solved by ``renormalizing''  the divergent masses and couplings to observable quantities. In this way the electroweak theory becomes a ``renormalizable'' theory, as proven by  't Hooft and   Veltman in the years 1971-1974 \cite{'tHooft:1972fi}. This worked well, as demonstrated by the excellent agreement between the calculated and observed radiative corrections, leading to correct predictions for the top quark - and Higgs boson mass from the electroweak precision experiments at the LEP collider at CERN.  't Hooft and   Veltman  were awarded the Nobel prize in 1999 after the confirmation of their calculations at LEP.

How does this contribution fit into this picture?  First I will discuss the electroweak precision experiments at LEP, which  tested the quantum structure of the SM in great detail.
A second topic has to do with physics beyond the SM. The SM is based on the product of the SU(3)xSU(2)xU(1) symmetry groups, so a natural question is: why three groups? And why can we not unify these groups into a larger group, like SU(5), having the SM  groups as subgroups \cite{Georgi:1974sy,Georgi:1980ic,Georgi:1974yf}?
The consequences are dramatic: since each SU(n) group is predicted to have n$^2$-1 gauge bosons, it doubles the number of gauge bosons (12 in the SM; 24 in SU(5)). In SU(5) the leptons from SU(2) and quarks from SU(3) are contained in the same multiplet, which leads automatically to  new lepton- and baryon number violating interactions between leptons and quarks. This inevitably leads to the proton decaying into leptons and quarks via the interactions with the new gauge bosons. In the standard SU(5) the proton lifetime was estimated to be of the order of $10^{31}$ years \cite{Georgi:1980ic}. The experimental limits\footnote{Since the background for proton decay experiments is provided by neutrinos, the discovery of  different backgrounds for up-going and down-going neutrinos led to the discovery of neutrino oscillations, which implies neutrino masses. This led to the Nobel prize for Koshiba in 2002.}  are two orders of magnitude above this prediction \cite{McGrew:1999nd,Abe:2014mwa}, thus  excluding   grand unification  in the SM, but not in the supersymmetric extension of the SM, which predicts a longer   proton lifetime  \cite{Marciano:1981un}.

To explain the long proton lifetime in a unified theory, the new  gauge bosons must be heavy.
How heavy? Presumably these gauge bosons get a mass by the breaking of the SU(5) symmetry into the SU(3)xSU(2)xU(1) symmetry, just like the W and Z bosons get a mass by breaking of the SU(2)xU(1) symmetry into the U(1) symmetry.
Above the  SU(5) breaking scale  one has a Grand Unified Theory (GUT) with a single gauge coupling constant.  Extrapolating  the precisely measured gauge couplings at LEP to high energies showed  that  unification is excluded in the SM, but in the supersymmetric extension of the SM the gauge couplings unify and interestingly, at a scale consistent with the long proton lifetime. This result, estimating simultaneously the GUT scale and the  scale of Supersymmetry  from a fit to the gauge couplings \cite{Amaldi:1991cn},  became quickly on the top-ten citation list and was discussed in widely read scientific journals \cite{Ross:1991qv,Hamilton:1991qu,Dimopoulos:1991au} and the daily press.

Supersymmetry was developed in the early 70's as a unique extension of the rotational and translational symmetries of the Poincar\'{e} group by a symmetry based on an internal quantum number, namely spin, see Ref. \cite{Ramond:2014qla} for a historical review and original references.  Supersymmetry requires an equal number of bosons and fermions, which can be realized  only, if every fermion (boson) in the SM gets a supersymmetric bosonic (fermionic) partner. This doubles the particle spectrum, but the supersymmetric partners have not been observed so far, so if they exist, they must be heavier than the SM particles.
 Not only gauge coupling  unification made Supersymmetry popular, since  it removes several shortcomings of the SM as well. Especially it  provides a dark matter candidate with the correct relic density  \cite{Jungman:1995df,Ade:2013zuv},  see e.g. Refs \cite{Haber:1984rc,deBoer:1994dg,Martin:1997ns,Kazakov:2010qn} for reviews.
 On the other hand, the main shortcoming of  Supersymmetry is the fact that none of the predicted supersymmetric partners of the SM particles have been observed, which could  be  a lack of luminosity or energy at the LHC, as will be discussed in the last section. And of course, other DM candidates exist as well \cite{Bertone:2010zza}.

\section{The Electron Positron Colliders}\label{s2}
After the discovery of  neutral currents in elastic neutrino-electron scattering in the Gargamelle Bubble Chamber, as discussed elsewhere in this volume, it was clear that a heavy neutral gauge boson must exist, as predicted by Weinberg \cite{Weinberg:1972tu}. The  weak gauge bosons were indeed observed  at CERN's proton-antiproton collider SPS, as discussed elsewhere in this volume. But it was clear, that precision experiments would need the clean environment of an \ee\ collider.  The CERN director, John Adams, who had just finished building the SPS, established in 1976  a study group to look into a Large Electron Positron Collider (LEP) for the production and study of the W- and Z bosons, predicted to have masses around 65 and 80 GeV. The group was led by Pierre Darriulat \cite{Darriulat:2004nf} and the famous Yellow Report was delivered half a year later \cite{Camilleri:1976mb}.  It contained  already many ideas on the physics potential and  first design ideas for  LEP, which was finally approved in 1982 and started taking data in 1989. The difficulties in realising such a large project has been described in the book entitled   `LEP: The Lord of the collider rings at CERN 1980-2000: The making, operation and legacy of the world's largest scientific instrument'' by Herwig Schopper, who was director-general at CERN during the construction of LEP. The book not only covers the technical, scientific, managerial and political aspects, but also discusses the sociological enterprises of building the large experimental collaborations of the LEP experiments with about 500 physicists per collaboration.  It also mentions the World-Wide-Web, which was invented during the LEP ope\-ration  by Berners-Lee and Cailliau in the IT department of CERN to improve the communication and data handling in the large LEP collaborations. 

During the same period SLAC set out to build a linear collider by equipping  the existing linear accelerator with damping rings and bending sections at the end to bring the sequentially accelerated  bunches of electrons and positrons into collision. Although on paper SLAC was expected to be ready  before LEP, the pioneering  task of colliding bunches of electrons and positrons in a linear collider took longer than anticipated, so finally, in the summer of 1989, the MARK-II collaboration observed its first  few hundred \Z\ events \cite{Abrams:1989ez} just before LEP came into operation.   

With a 45 kHz bunch crossing rate at LEP versus a 120 Hz repetition rate at the SLC the data sample at LEP  quickly outgrew the one at the SLC, since at its peak luminosity of    $10^{32}$ cm$^{-2}$s$^{-1}$ each LEP experiment collected about 1000 \Z\ bosons per hour. A brief review of all the ups and downs on the way to reach a luminosity at LEP above its design value was given at the Topical Seminar on `The legacy of LEP and SLC''  in    Sienna in 2001 \cite{assmann}.  
This review on the LEP accelerator describes also the precise beam energy determination via spin depolarisation techniques, which can determine the beam energy to 0.2 MeV or a relative accuracy of $5\cdot10^{-6}$. In addition, the many surprises, like the correlation of the tides from the gravitational interaction between the moon and the earth or the amount of water in Lake Geneva with the beam energy, are described. These effects of a few MeV in the beam energy correspond to a change in the orbit length of a few mm, caused by the elasticity of the earth's crust. Also the short term energy fluctuations  from the fast TGV train between Geneva and Paris, for which the LEP magnets turned out be a good current return path, were finally understood after these fluctuations were absent during a railway strike in France. 
\begin{figure}[]
\begin{center}
\includegraphics[width=0.45\textwidth]{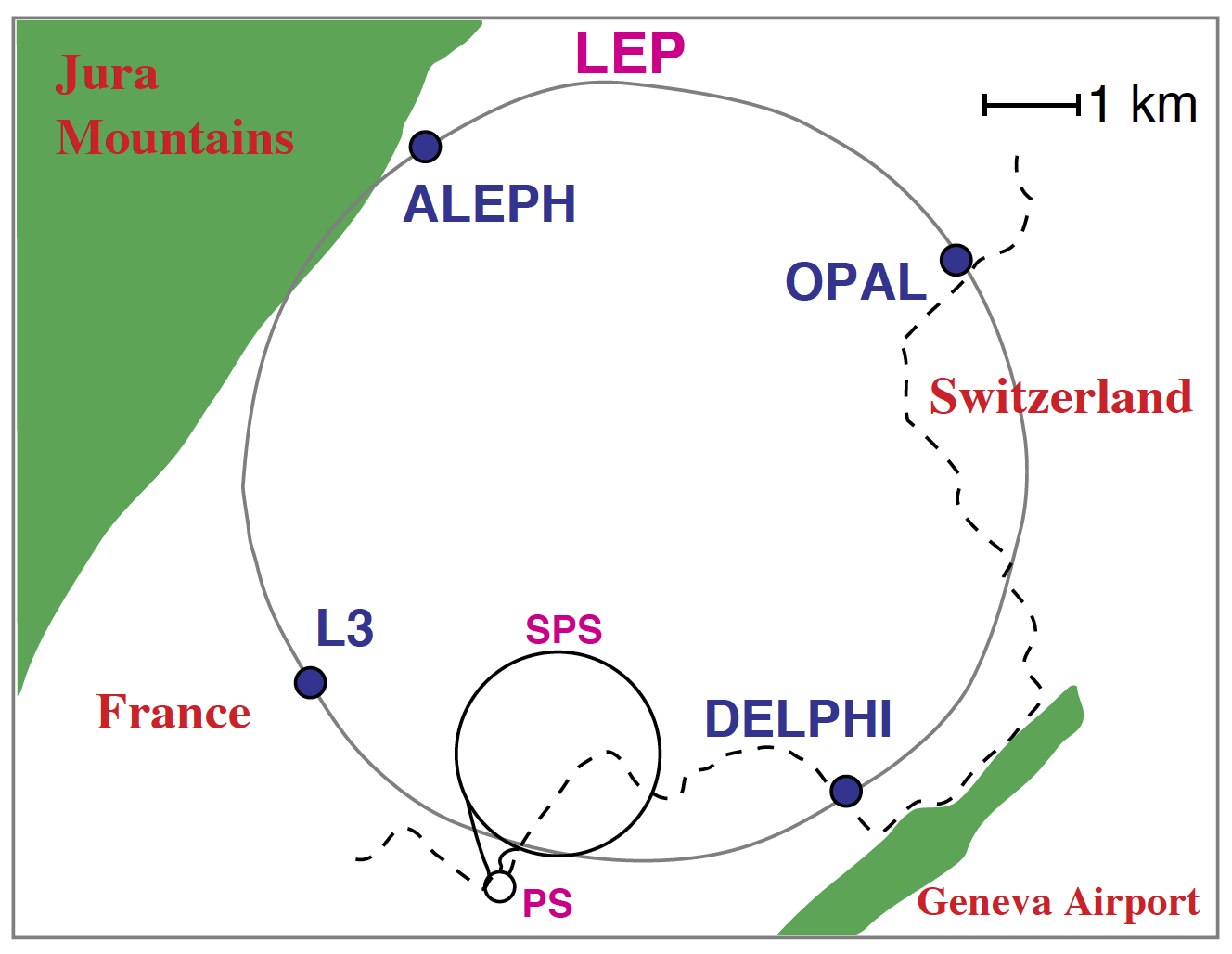}\hspace*{6mm}
\includegraphics[width=0.48\textwidth]{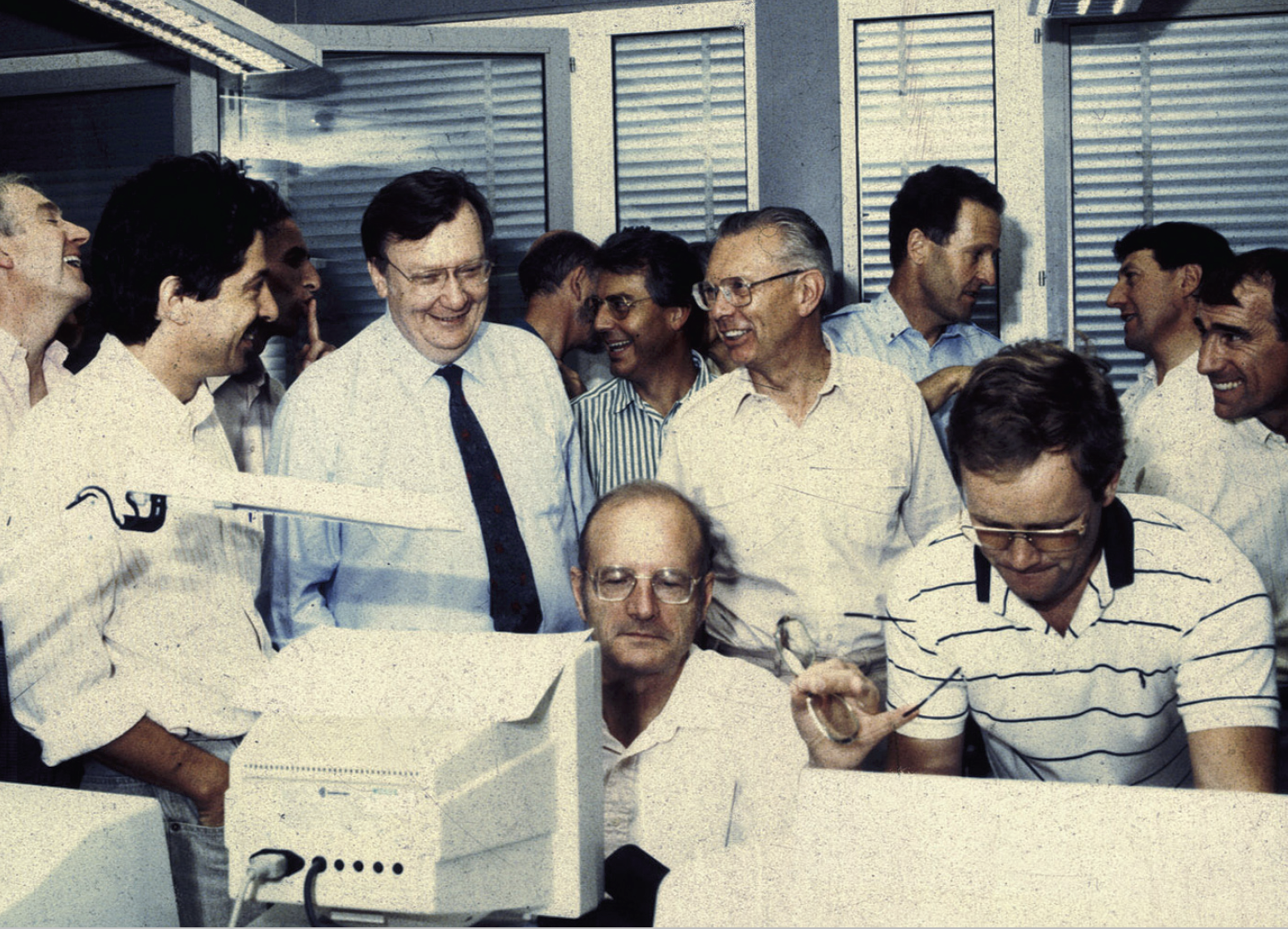}
\center{\hspace*{25mm}(a)\hspace*{0.5\textwidth}(b)}
\caption[The LEP storage ring] {(a): The LEP storage ring with the four experiments and its pre-accelerators (PS and SPS). (b): Happy faces during the start of LEP in July 1989.}
\label{f1}
\end{center}
\end{figure}
The final uncertainty  of about 2  MeV in the \Z\ mass  from the beam energy is  considerably larger, mainly because  the field of the dipole magnets varies with time.
A schematic picture of the 27 km long LEP tunnel and its experiments is shown in Fig. \ref{f1}, together with the joyful faces after the start of the operation in July, 1989. 

After LEP started running  the SLC made an amazing improvement by providing highly polarized beams, which are a sensitive probe of the weak interactions, in which left- and right-handed particles have different couplings. These data were largely collected by the SLD detector, which could determine the electroweak mixing angle with comparable  precision in spite of the much smaller data sample of about half a million  \Z\ bosons  (in comparison with 17 million events for the combined LEP experiments).
At LEP a polarization scheme had been studied in great detail as well \cite{Blondel:1987wr}, but  finally it was discarded in favour of going to higher energies as quickly as possible.

 In 1995 LEP was upgraded to reach the WW and ZZ  pair production threshold  and later on  up to 208 GeV  (by adding more accelerating cavities) in the hunt for the Higgs. One could set a 95\% C.L. lower limit of 114.4 GeV on the Higgs mass \cite{Barate:2003sz}, just 11 GeV short of the Higgs mass found  at the LHC in 2012. This higher energy could have been reached, if all available space at LEP would have been filled with superconducting cavities, in which case Higgs masses up to the SUSY upper limit of 130 GeV \cite{Djouadi:2005gj} could have been reached, see e.g. the   review  on LEP and SLC results  \cite{Treille:2002iu}.   However,  the time  and  financial pressure from the LHC in competition with a Tevatron upgrade (the SCC had been abandoned two years before  in 1993 due to budget problems) led to the decision to stop   LEP operation in 2000. Of course, in retrospect, the Higgs boson could have been discovered 10 years earlier at LEP and studied in the clean environment of an \ee collider.

\section{The four LEP detectors}\label{s3}
In total four LEP detectors were approved: ALEPH (Appartus for LEP Physics) \cite{Decamp:1990jra}, DELPHI (Detector with Lepton and Hadron Identification) \cite{Aarnio:1990vx}, L3 (Letter of Intent 3) \cite{Adriani:1993gk} and OPAL (Omnipurpose Apparatus for LEP) \cite{Ahmet:1990eg}. 
  All detectors are  large 4$\pi$ detectors with  sizes of  typically 10 m in each direction and a weight of  up to thousand medium-sized cars. They are  designed to study the hadronic, electromagnetic and leptonic components of the final states of the \Z\ boson, but they differ in experimental techniques, like resolution of the magnetic spectrometers, the electromagnetic- and hadronic calorimeters and the extent of particle identification. In addition, all detectors were upgraded  to have silicon based vertex detectors just outside the beam pipe (see Ref. \cite{Hartmann:2009zza} for a review), which allowed to locate the primary collision vertex typically with a precision of a few $\mu m$. This allowed to tag jets from b- and c-quarks by their secondary vertex, since the long-lived B- and D-mesons travel on average several mm before decaying and producing a secondary vertex.   

  The resources and manpower needed for  large detectors require large collaborations, typically 250 at the start of LEP and climbing to 500 physicists at the end. Around 20-50 institutions are involved, most of them from the European member states, but also from Asia, Isreal,  Russia and the US.  The ALEPH and DELPHI detectors were considered ``risky'' by the LEP Experiments Committee, since they used superconducting magnets\footnote{The DELPHI solenoid was with 6.2 m in diameter, 7.2 m in length and a field of 1.2 T the world's largest superconducting magnet.} and time-projection-chambers as 3D tracking devices. In addition, ALEPH used large liquid argon electromagnetic calorimeters, while DELPHI applied the 3D time-projection idea also to the electromagnetic calorimeter and installed  in addition  Ring Imaging Cherenkov (RICH) detectors for hadron identification. The L3 and OPAL detectors used more conventional techniques, like  wire chambers for tracking, a warm magnet and scintillating crystals as electromagnetic calorimeters. 

   One may wonder why one needed as many as four experiments at LEP. Would two not have been enough?  The four detectors do not only provide redundancy, but have different systematic uncertainties. The redundancy turned out to be of utmost importance to investigate fluctuations, like the many standard deviations excess in 4-jets \cite{Buskulic:1996hx}   and the Higgs-like signal with a mass around 115 GeV \cite{Heister:2001kr}.   If the Higgs-like signal, mainly based on three ALEPH events, was combined with all other experiments, the significance was less than 2$\sigma$. We now know from the observed Higgs mass that it was indeed a statistical fluctuation. Also the 4-jet excess turned out to be a fluctuation, as was clear from the combined data of all experiments \cite{Abreu:1998ih}.

And last, but not least, in spite of the impressive data sample, in ratios involving leptonic decay modes,  the statistical errors still dominate, so they profit from a factor two lower error after combining the data from the four experiments. The combination holds also the risk of dominating common systematic theory errors, which, if not correctly estimated, may change the results. We will see examples in the discussion of the coupling constants.

In spite of being competitors the four experiments collaborated in working groups to combine all experimental data in order to get the most precise answers to the questions asked. Prominent working groups were the Electroweak Working Group (EWWG), the Heavy Flavor Working Group, the Higgs Working Group and the Working Group on searches. This working in large collaborations  and even combining data from different collaborations was a turning point in the history of high energy physics, not only important for LEP, but also a sociological exercise for LHC, where the largest collaborations grew to about 3000 collaborators.
\section{Renormalizing the infinities of the SM}\label{s4x}
{\it ''A confrontation with infinity''} was the title of 't Hooft's Nobel Lecture. He showed that the high energy divergences  in the SM can be tamed by  renormalization techniques. Here one is subtracting in principle infinitely large numbers. The resulting answer is still reliable, as was proven by the electroweak precision measurements at LEP. To appreciate the significance of this result,  I shortly describe the physical ideas behind the renormalization of masses and couplings. Take e.g. an electron. If it is a point-like particle, the potential energy in the electric field goes to infinity, if the distance goes to zero. In quantum language this would mean an infinite amount of photons. On short time scales they fluctuate into an infinite amount of e.g. \ee pairs. The field energy and its \ee pairs increase the mass of the electron, since energy implies mass according to Einstein’s $E=mc^2$. So the total mass can be written as the sum of two contributions: $m=m_{bare}+\Delta m$, where $m_{bare}$ is the bare mass and $\Delta m$ the contribution from the surrounding field, which becomes infinite as $R\rightarrow 0$. To get the mass in agreement with the observed mass, one has to let the bare mass go to minus infinity, i.e. one renormalizes the calculated mass to the observed mass. But this is not all: the virtual particles will orient themselves in the electric field, which causes a decrease of the electric field, just like the polarization of a dielectric material inside a capacitor decreases the electric field. This screening of the bare charge by the  "vacuum polarization", leads to an energy dependence of the couping constants, since at high energies one looks at distances close to the bare charge, unshielded by the vacuum polarization.The calculated charge can again be renormalized to the observed charge at large distances, e.g. the fine structure constant in atomic physics or Thomson scattering at zero momentum transfer.  The energy dependence of the coupling constant can be calculated from the loop diagrams of photons fluctuating into \ee\ pairs,  which can be described by the Renormalization Group Equations (RGE)\cite{Wilson:1973jj}. 
A heuristic derivation of the RGEs was given by 't Hooft's  in his Nobel Lecture \cite{nobelall}.
There he compares the use of differential equations in classical mechanics, like the movement of planets with the use of differential equations in quantum mechanics (QM). In classical mechanics one can calculate the speed of an object by $v= \Delta x/ \Delta t$ or as a differential equation  $ v=dx/dt$  for small distances and times to get the  most accurate speed. There is no problem with this.  However, in QM one has to take into account the quantum fluctuations at small distances. One can check the precision of the solution  by choosing different scales $\Delta x$  in $v(x)= \Delta x/ \Delta t$ assuming one knows $v(x)$. 
\begin{figure}
\begin{center}
\begin{minipage}{0.4\textwidth}
\includegraphics[width=0.8\textwidth]{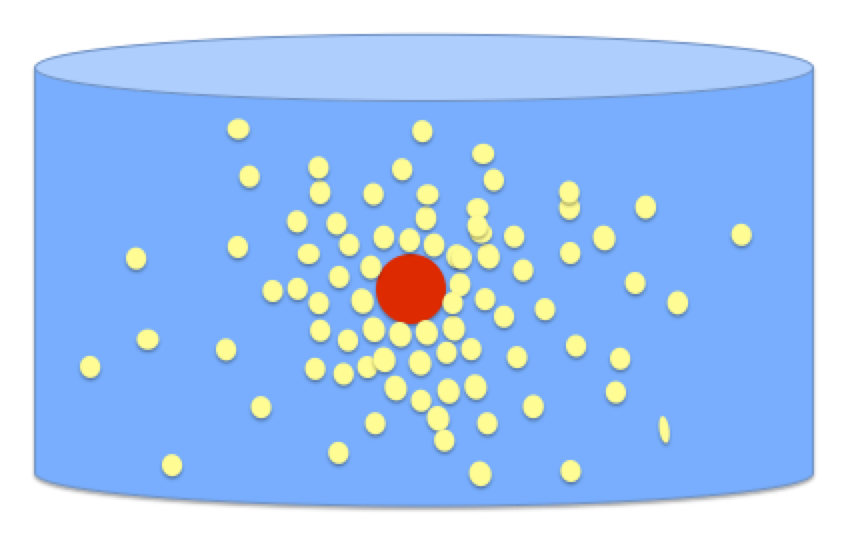}
\center(a)
\end{minipage}\hspace*{5mm}
\begin{minipage}{0.4\textwidth}
\includegraphics[width=0.8\textwidth]{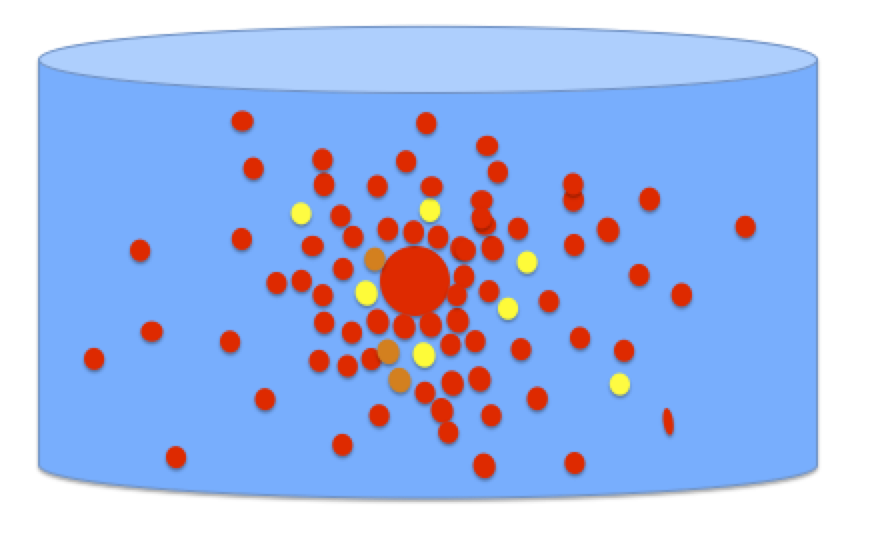}
\center(b)
\end{minipage}
\caption{Charge distribution around an electric charge (a)  and a color charge (b). The light (yellow) dots indicate fermions (lepton and quark pairs in the electric field, quarks pairs in the color field) , the dark (red) dots the gluons. Since the fermion pairs have an opposite charge, they orient themselves in the field and screen the "bare" charge in the center. The gluons enhance the "bare" charge in the center. As a consequence, the effective coupling at high energy is increased (reduced) in QED (QCD). }
\label{f2x}
\end{center}
\end{figure}
E.g. if we introduce a scale parametrized by a parameter $1/\mu$ (where $\mu$ would be the energy in a quantum field theory) we find for $dx\equiv d(1/\mu)=-d\mu/\mu^2$ or $d\mu/\mu=-d x/x$: $\frac{1}{\mu} \frac{d\mu}{dx}=f(x)$.
One can write a similar equation for the energy dependence of the coupling constant $\lambda$ (e.g. the quartic coupling in the Higgs potential or $g^2$ in QED):
\begin{equation}  
 \frac{1}{\mu} \frac{d\mu}{d \lambda}= \beta(\lambda).
    \label{beta}
\end{equation}
\noindent
The slope of the logarithm of the energy dependence of the coupling constant is given by the beta function $\beta(\lambda)$. A positive (negative) beta function means that the coupling  increases (decreases) with energy. For QED the beta function is positive and the fine structure constant changes from 1/137.035999074 at low energy to 1/127.940 at LEP I energies. For QCD the beta function is negative, so the strong coupling constant decreases with energy. 
 This decrease   is the origin of asymptotic freedom, for which  Politzer, Gross  and Wilczek  got the Nobel Prize in 2004. The physics behind asymptotic freedom is the fact that the gluons carry themselves color, so the gluons interact with themselves. This leads to an anti-screening of the bare color charge, so if the anti-screening is stronger than the screening by the quark pairs in the colour field, the interactions at high energy see only the bare colour charge, which is small and  quarks are ''free'' at high energies. So  the bare charge around an electron and quark is screened by fermion-antifermion pairs (light (yellow) in Fig. \ref{f2x}), but enhanced by gluons  (dark (red)).  Since the gluon self-interaction prevents gluons to move off to infinity, the field lines between two colour charges are confined to a string between them in contrast to the electric field, in which case the field lines spread out to infinity.
The energy density in the  string becomes so high  at large distances, that it is energetically favorable to convert the energy into mass by creating quark-antiquark pairs, which leads to the observation of  jets of particles  instead of bare quarks. The three jet events in \ee\  annihilation at PETRA in 1979 heralded the discovery of the gluon as a real physical entity, produced by gluon radiation of  quarks, see Ref. \cite{Soding:2010zz} for a review.
\section{Quantum corrections to the W- and Z boson masses}\label{s5}
The interaction between two matter particles can be mediated by a gauge boson, which leads for massless gauge bosons to a propagator factor  $g^{\mu\nu}/q^2$ in the Feynman diagram, where  $q$ is the momentum flowing through the propagator and  $g^{\mu\nu}$ is the Minkowski metric with Lorentz indices $\mu$ and $\nu$. For a massive gauge boson with mass $m$ the propagator gets an additional factor $k_\mu k_\nu/m^2$. This factor, originating from the longitudinal spin degree of freedom of the gauge boson,  becomes infinite, if the momenta $k$ of the incoming and outgoing particles become infinite.  This infinity can only be compensated by adding a counterpiece, so in general the propagator of a massive particle is:
\begin{equation}  
 \frac{g_{\mu\nu} - \frac{k_\mu k_\nu}{m^2}}{q^2-m^2+i\epsilon}+\frac{\frac{k_\mu k_\nu}{m^2}}{q^2-\frac{m^2}{\lambda}+i\epsilon},
    \label{prop}
\end{equation}
where the gauge parameter $\lambda$ can be chosen as 0, 1 or infinity, which corresponds to the unitarity gauge, Feynman or 't Hooft gauge and Landau or Lorentz gauge, respectively. The last term in Eq. \ref{prop} represents the propagator of a scalar particle for $\lambda=1$, i.e. in the Feynman or 't Hooft gauge.   In this case the physics behind the compensation of the  $k_\mu k_\nu/m^2$ term is simple: the infinity in the amplitude of longitudinal W boson exchange is  compensated  by the exchange of a Higgs boson, so the calculated cross section will not pass the unitarity limit\footnote{Weinberg noted in his Nobel  Lecture\cite{nobelall}, that he did not succeed in proving the renormalizabi\-lity, since he was using the unitarity gauge, which has the advantage of exhibiting the true particle spectrum, but the disadvantage of  obscuring the renormalizability, as is obvious from Eq. \ref{prop}.}. As 't Hooft noted in his Nobel lecture\cite{nobelall}: people knew that gauge boson masses can be generated by the Higgs mechanism, but they did not know that this was a {\it unique} solution, since at the same time it removes the infinities, thus making the theory renormalizable\footnote{'t Hooft noted also that the  unitarity problem did not bother him, since he had discovered already that the SU(3) group had a negative $\beta$-function, thus decreasing the cross section at high energy. However, he did not realize ``what treasure he had here'', so he did not connect it to asymptotic freedom. He  expected anyway that all experts would know about the different signs of the $\beta$-function in QED and QCD}. 
\begin{figure}[t]
\begin{center}
\begin{minipage}{0.65\textwidth}
\includegraphics[width=\textwidth]{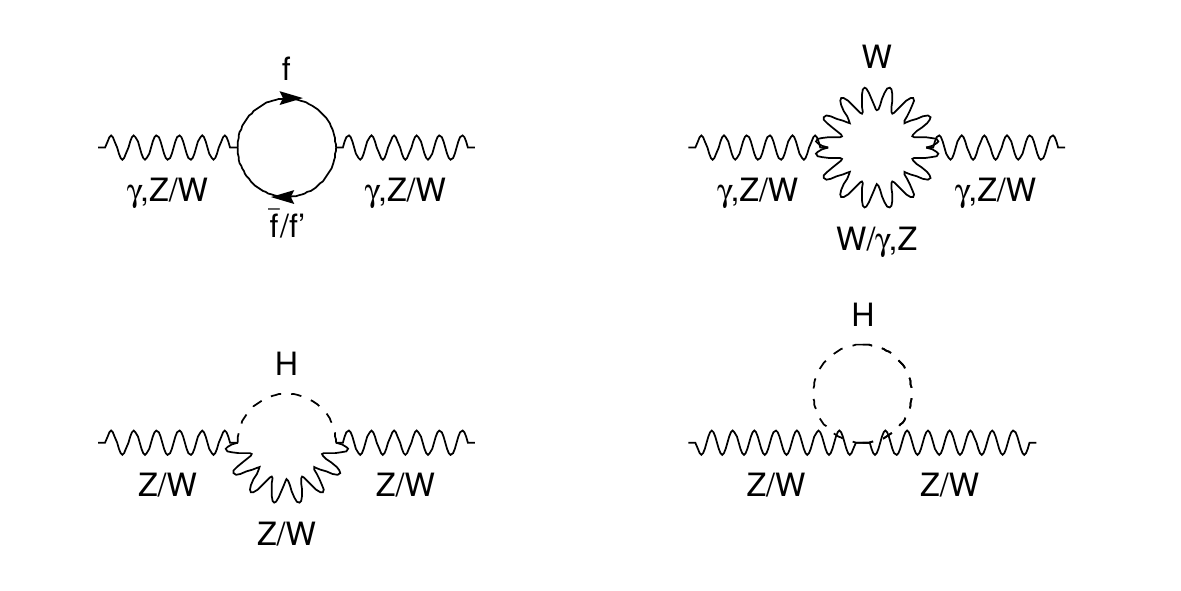}
\end{minipage}
\begin{minipage}{0.34\textwidth} 
\includegraphics[width=\textwidth]{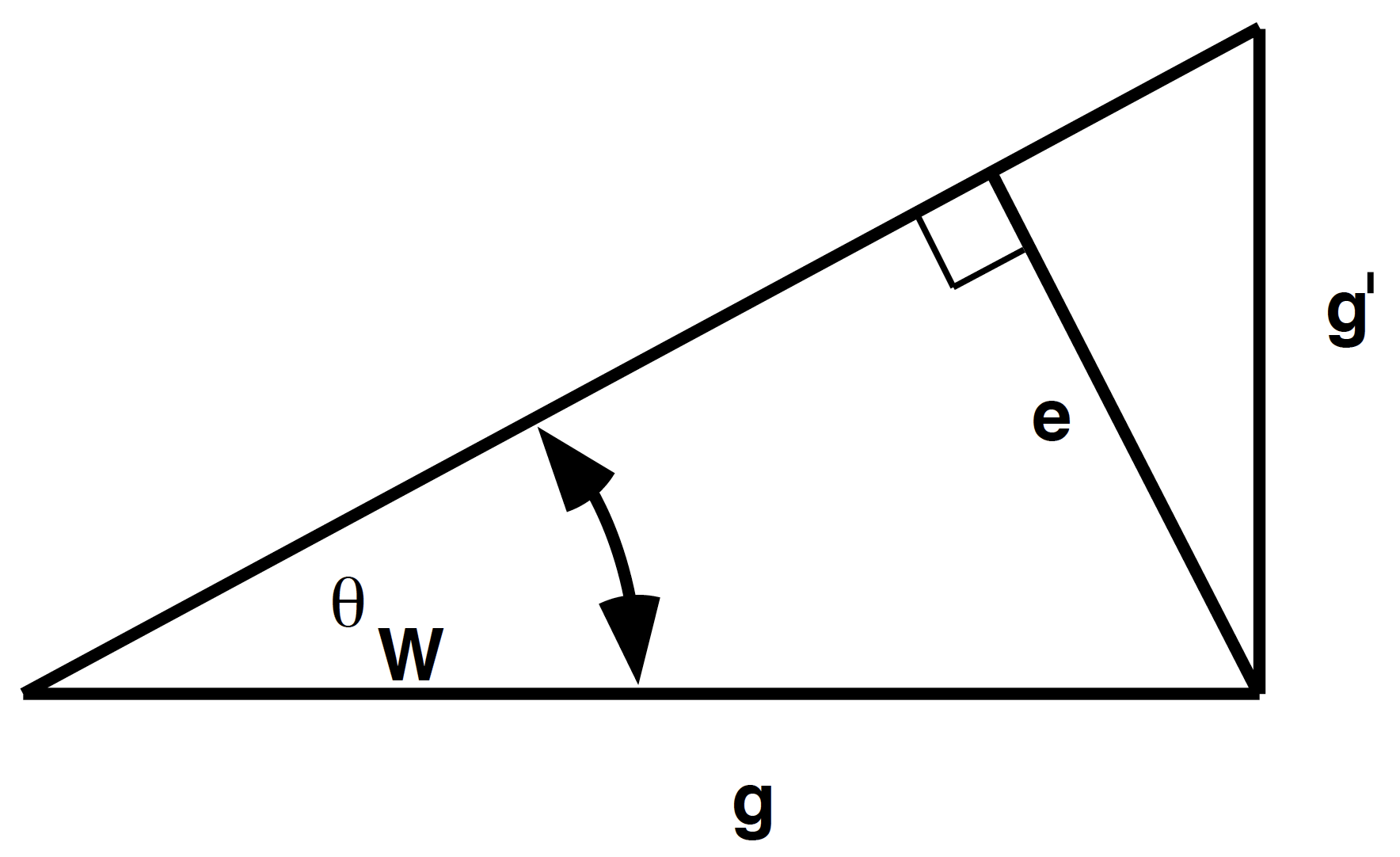}\vspace*{-3mm}
\end{minipage}
\center{\hspace*{25mm}(a)\hspace*{0.5\textwidth}(b)}
\caption{(a): Loop corrections to the SM propagators. (b): Relations between gauge couplings.}
\label{f2}
\end{center}
\end{figure}
An important aspect of proving the renormalizabi\-lity of the SM is a recipe how to handle technically the divergences. This was done most conveniently by dimensional regularisation, as discussed by 't Hooft and Veltman \cite{'tHooft:1972fi}. But what was of utmost importance for LEP: with such a renormalization scheme Veltman could calculate the radiative corrections from  Higgs boson - and fermion loops to the weak gauge bosons, depicted in Fig \ref{f2}a, and found surprisingly that the corrections depend  quadra\-tically on the top mass \cite{Veltman:1980fk}.  For the Higgs mass the quadratic term happens to have zero amplitude\footnote{Veltman called this the ``screening theorem'', since the Higgs boson ``screens'' itself against detection via observable radiative corrections.},  so only a logarithmic dependence is left. 
After electroweak symmetry breaking via the Higgs mechanism the mass eigenstates become linear combinations of the gauge bosons of the original (symmetric) Lagrangian ($W^i,  i=1,2,3$ for SU(2) and B for U(1)): $W^\pm=(W^1\mp W^2)/\sqrt{2}$, $Z=-B\cos\theta_W +W^3 \sin\theta_W$, $\gamma=B\sin\theta_W+W^3\cos\theta_W$, where
the electroweak  mixing angle $\theta_W$ 
is determined by the ratio of the coupling constants of the U(1) and SU(2) groups: $\tan\theta_W=g^\prime/g$ and its relation to the electric charge is depicted in Fig. \ref{f2}b, implying $e=g\sin\theta_W$.

Since  the Higgs mechanism predicts the gauge boson masses to be proportional to the gauge couplings one finds:
\begin{equation}
\label{eq:rho}
 \cos\theta_W  =
 \frac{g}{\sqrt{g^{\prime 2}+g^2}} =\frac{M_W}{M_Z} \hspace*{5mm} {\rm or } \hspace*{5mm} \rho_0=\frac{\MW^2}{\MZ^2\cos^2\theta_W}
\end{equation}
In the SM $\rho_0=1$, but it can deviate from 1 for a more complicated Higgs structure.
The  muon decay proceeds via W exchange, so the W mass is related to the  muon decay constant:
${G_F}={\pi\alpha}/({\sqrt{2}\swsq M_W^2})$, which leads to
$M_W^2 =A^2/{\swsq}$, 
$M_Z^2 = {A^2} /({\swsq\cwsq})$ with
 $A=\sqrt{{\pi \alpha}/{\sqrt{2}G_F}}=37.2805$ GeV.  This  value of $A$ leads with  $\swsq=0.2314$  to  $\MZ$=88 GeV.
However, these relations hold only at tree level and  are modified by loop corrections (see Fig. \ref{f2}a):
\begin{align}
\label{sw}
 \sinw=\left(1-\frac{\MW^2}{\MZ^2}\right)  = \frac{A^2}{1-\Delta r} \, , \
\end{align}
where the radiative corrections have been lumped into $\Delta r$, which depends quadra\-tically on the top mass and logarithmically on the Higgs mass.  These definitions are valid in the so-called on-shell renormalization scheme \cite{Sirlin:1980nh,Kennedy:1988sn,Bardin:1989di,Hollik:1988ii}, in which case the electroweak mixing angle is defined by the on-shell masses of the gauge bosons: $\sinw\equiv 1-\MW^2/\MZ^2$. In this scheme $\Delta r \approx \Delta r_0 - \rho_t/\tan^2\theta_W$, where $\Delta r_0 = 1-\alpha/\alpha(\MZ)=0.06637(11)$ and $\rho_t=3\GF\mt^2/8\sqrt{2}\pi 2 = 0.00940 (\mt/173.24 GeV)^2$. The latter term shows the quadratic top quark dependence, which is enhanced by $1/\tan^2\theta_W=3.32$, so the negative $\mt$ corrections  are almost 50\% of the dominant  $\Delta r_0 $ correction. 

The on-shell renormalization scheme has been  used by the EWWG for the ana\-lysis of the LEP electroweak precision data. An alternative scheme,  the modified minimal subtraction $\MSbar$ scheme \cite{Fanchiotti:1992tu}, is extensively used in QCD. In this scheme  the electroweak mixing angle is not defined by the masses ($\sinw\equiv 1-\MW^2/\MZ^2$), but defined by the tree level values of the couplings: $\sin\theta_{\MSbar}\equiv g^\prime / \sqrt{g^{\prime 2}+g^2}$  (see Fig. \ref{f2}b) with all couplings defined at the Z mass \footnote{The values of the electroweak mixing angles  are related  in both schemes by 
$\sin^2\theta_{\MSbar}= c(\mt,\MH)\sinw=1.0344\pm 0.0004)\sinw$, where  $c(\mt,\MH) = 1+\rho_t$,   so in this case  the couplings become dependent on the top mass.}. 
 The total cross section must be independent of such a choice, so the masses in the $\MSbar$ scheme must be redefined to:
$\MW^2=A^2/ (\sin^2\theta_{\MSbar} (1-\Delta r_{\MSbar}))$ and $\MZ^2=\MW^2/(\rho_{\MSbar} \cos^2\theta_{\MSbar})$, where $\Delta r_{\MSbar}\approx \Delta r_0$ and  $\rho_{\MSbar} \approx 1+\rho_t$. With these definitions $\MW$ becomes practically independent of the top mass. This is reasonable, since its value is determined by $\GF$,  which has the radiative corrections  absorbed in the measurement.   All top mass dependent corrections are now included in $\MZ$ and the couplings between the Z boson and the fermions. 

The W  bosons couple only to left-handed particles and right-handed antiparticles with a strength given by the weak charge $I_3$, which is +1/2 for the neutrinos and up-type quarks, -1/2 for the charged leptons and down-type quarks. The right-handed particles have vanishing weak charge, i.e. $I_3 \approx 0$ \footnote{The  difference in the weak charge between left and right is the basis for the famous parity violation, observed in 1954 by C.S. Wu and explained by Yang and Lee, who received for this fundamental discovery the Nobel prize in 1957\cite{nobelall}.}.  
The photon couples equally to left and right-handed particles, so after mixing of $W^3$ and $B$  the  Z couplings  obtain an electromagnetic component $-Q_f\, \sin^2\theta_{W}$: 
$g_L^f=  \sqrt{\rho_f}(I_3^f-Q_f\,\sin^2\theta_{W})$ and $g_R^f= -Q_f\,\sin^2\theta_{W}$.
The vector and axial vector couplings are defined  as:
\begin{equation}
\label{couplings}
  g_V^f  =   g_L^f+g_R^f=    \sqrt{\rho_f} (I_3^f-2Q_f\,\sin^2\theta_{eff})  \hspace*{1.2cm}
  g_A^f  =    g_L^f-g_R^f=     \sqrt{\rho_f} I_3^f,     
 \end{equation}
where $\sin^2\theta_{eff}^f=\kappa^f\sinw$ is the effective mixing angle, i.e. the one including radiative corrections.   At tree level $\rho_f=\rho_0=1$,
 except for the b quark, since the vertex correction from a triangle loop with  top quarks and a W boson changes slightly the b quark production cross section.
In this case \cite{Agashe:2014kda} 
\begin{equation}\label{rhob}\rho_b\approx 1+\frac{4}{3}\rho_t  \hspace*{1cm} {\rm and} \hspace*{1cm} \kappa_b \approx 1+\frac{2}{3}\rho_t. \end{equation} The difference between the effective mixing angle and the  $\MSbar$ mixing angle for $f\ne b$ is small and almost independent of the Higgs and top mass: $\sin^2\theta_{eff}^f -  \sin^2\theta_{\MSbar}^f =  0.00029$\cite{Agashe:2014kda}, 
\begin{figure}[]
\begin{center}
\begin{minipage}{0.47\textwidth}
\includegraphics[width=1.02\textwidth,height=0.96\textwidth]{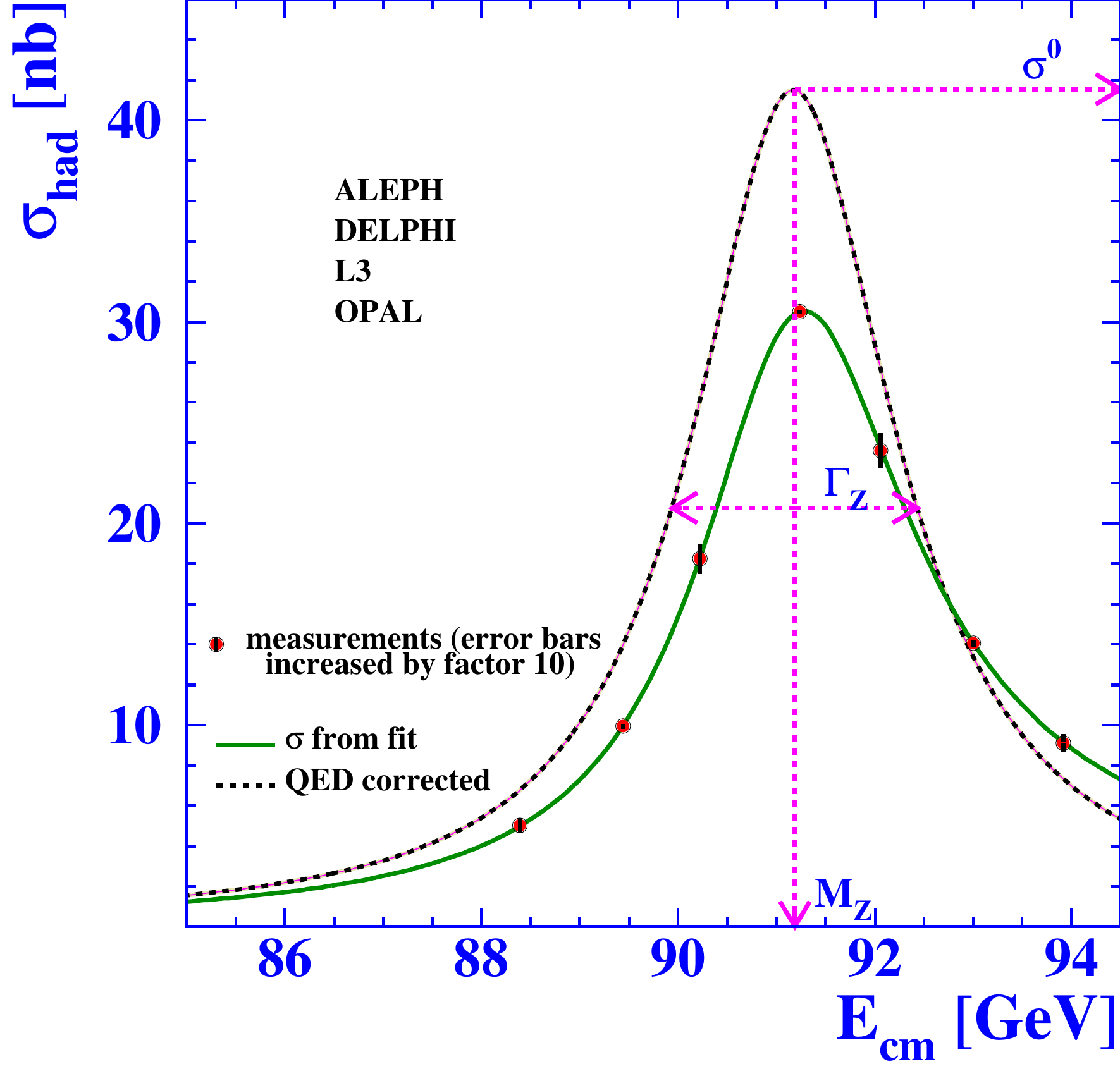}\vspace*{-2mm}
\end{minipage}\hspace*{7mm}
\begin{minipage}{0.47\textwidth} 
\includegraphics[width=1.02\textwidth,height=1.06\textwidth]{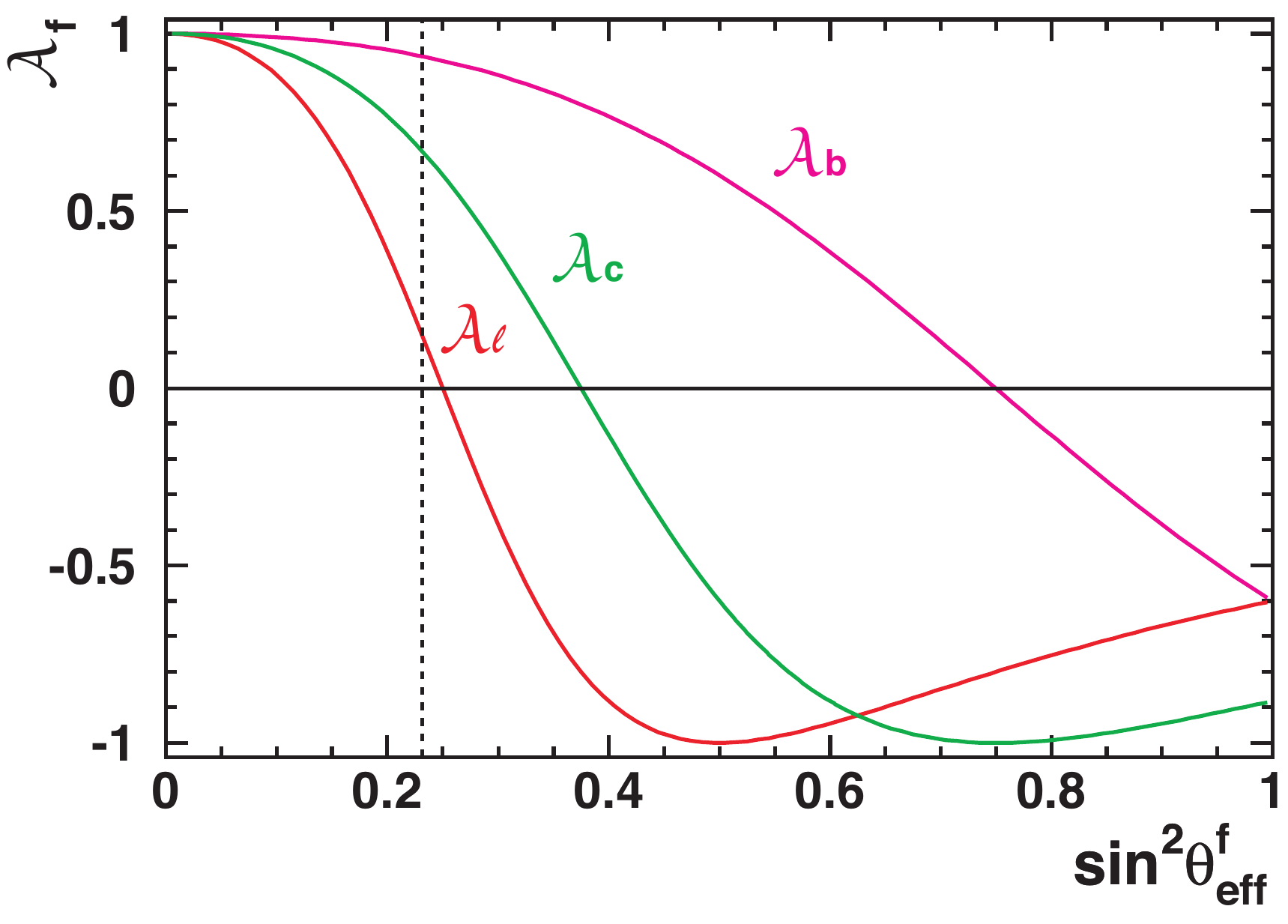}\vspace*{-6mm}
\end{minipage}
\center{\hspace*{10mm}(a)\hspace*{0.5\textwidth}(b)}
\caption{(a): Hadronic cross section with and without radiation. (b): Sensitivity of the asymmetry to \sinw\ for various fermionic final states. From Ref. \cite{ALEPH:2005ab}.}
\label{f3}
\end{center}
\end{figure}
an important relation, since the LEP electroweak working group always determines  $\sin^2\theta_{eff}^l$, but for gauge coupling unification one needs the value in the $\MSbar$ scheme.
\section{SM cross sections, asymmetries and branching ratios}\label{s6}
The differential cross section for \ee\ annihilation into fermion pairs can be written as \cite{Agashe:2014kda}:
\begin{equation} \label{eq:cross}
\frac{2s}{\pi}\frac{1}{N_c^{\rm f}}\frac{d\sigma_{\rm ew}}{d\cost}(\eeff)\ = \ 
\alpha^2(s)\left[F_1(1+\cos^2\theta)+2F_2\cos\theta\right]+B,
\end{equation}
where $F_1=Q_e^2Q_f^2 \chi Q_e Q_f g_V^e g_V^f\cos\delta_R + \chi^2(g_V^{e2}+g_A^{e2})(g_V^{f2}+g_A^{f2})$, $F_2=-2\chi Q_e Q_f g_A^e g_A^f\cos\delta_R +4\chi^2 g_V^e g_A^e g_V^f g_A^f$, $\tan\delta_R=\MZ\GZ/(\MZ^2-s)$, $\chi(s)=  \left({\GF s\MZ^{2}}\right)/\left({{2\sqrt{2}\pi\alpha(s)}\left[(s-\MZ^{2})^2 + \GZ^2\MZ^2\right]^{1/2}}\right)$,
 $\alpha(s)$ is the energy dependent electromagnetic coupling and $\theta$ is the scattering angle of the out-going fermion with respect to the direction of the e$^-$ beam. The color factor $N_c^{\rm
f}$ is either one (for leptons) or  three (for quarks), and $\chi(s)$ is the
propagator term; B  represents small  contributions from the electroweak box graphs. 
The cross section is asymmetric around the peak,  as illustrated in Fig. \ref{f3}a.: at energies above the peak the cross section is higher, because of QED corrections, mainly from single photon radiation off the incoming beams. After radiating a photon the effective CM energy is reduced, thus increasing the cross section at the effective CM energy. The asymmetry in the cross section can be described by a radiator function \cite{Bardin:1999ak}, which is usually taken into account in the fitting function.
%

Since an axial vector changes its sign in a mirror, the axial vector coupling is responsible for the  cosine term in Eq. \ref{eq:cross}, which leads to asymmetries in the angular dependence of the cross section or in the polarization asymmetry in  case of  polarized beams.
Defining  for a fermion $f$: 
\begin{align}
A_f = \frac{2 g_V^f g_A^f }{g_V^{f2}+g_A^{f2}}=\frac{2g_A^f/g_V^f}{1+(g_A^f/g_V^f)^2},
\end{align} one finds for the  forward--backward asymmetries $A_{FB}$
from the cross sections integrated over the forward ($\sigma_F$)
and the backward ($\sigma_B$) hemisphere,
 $A_{FB}  = ({\sigma_F-\sigma_B})/({\sigma_F+\sigma_B}) =  3 A_e/4 A_f $
and the left--right asymmetry from the cross sections $\sigma_{L,R}$
for left- and right-handed polarized electrons,
$ A_{LR}  = ({\sigma_L-\sigma_R})/({\sigma_L+\sigma_R})  = A_e $,
all of them being determined by the ratio $g_A/g_V$, so they are sensitive to the electroweak mixing angle $\sinw$ (see Eq. \ref{couplings}), especially for the leptons, since $g_V$ changes sign for $\sinw=1/4$, while for quarks the zero-crossing happens at much larger values, as shown in Fig. \ref{f3}b. However, for quarks the asymmetries are larger for $\sinw=1/4$, thus reducing the relative systematic errors. 
The weak mixing angle   completely determines the branching fractions $\sum(g_V^2+g_A^2)/\sum_{tot}$, where the numerator is summed over the fermions considered and $\sum_{tot}$ is the sum over all possible fermions. The branching fractions, calculated for $x=\swsq=0.2315$, agree reasonably well with observations, as demonstrated in Table \ref{t1}. The small discrepancies with the observed values originate from neglected  fermion masses and mssing higher order radiative corrections, since only the dominant radiative correction at the b-vertex from the top loop  (Eq. \ref{rhob}) has been taken into account.
\begin{table}
\tbl{Z branching ratios for $x=\swsq=0.2315$.}
{\begin{tabular}{|c|c|c|c|c|c|}\hline
\multicolumn{1}{ |c| }{Particles } &
\multicolumn{3}{|c|}{Couplings (Eq. \ref{couplings})}  &
\multicolumn{2}{ |c| }{   Branching ratios  }  \\ \hline
\multicolumn{1}{|c| }{Symbol} &
\multicolumn{1}{|c|}{$g_V$} &
\multicolumn{1}{|c|}{$g_A$}&
\multicolumn{1}{|c|}{$\sum( g_V^2+g_A^2)$}&
\multicolumn{1}{|c|}{calc.}  &
\multicolumn{1}{|c| }{obs.} \\ \hline
\hbox{\vrule height 9pt depth 4.5pt width 0pt}
$\nu_e,\nu_\mu,\nu_\tau$&\mathstrut$\frac{1}{2}$ &	$\frac{1}{2}$&	3($\frac{1}{2}$ )$^2$ +3($\frac{1}{2}$ )$^2$ &	20.5\%	&20.00$\pm$ 0.06\%\\ \hline
\hbox{\vrule height 9pt depth 4.5pt width 0pt}
$e,\mu,\tau$&\mathstrut$-\frac{1}{2}$ + 2x&	$-\frac{1}{2}$&	3($-\frac{1}{2}$ + 2x)$^2$ + $3(\frac{1}{2})^2$&	10.3\%	&10.097$\pm$ 0.0069\%\\ \hline
\hbox{\vrule height 9pt depth 4.5pt width 0pt}
u,c&\mathstrut$\frac{1}{2}$ - $\frac{4}{3}$x &	$\frac{1}{2}$&	6($\frac{1}{2}$  - $\frac{4}{3}$x)$^2$ + 6($\frac{1}{2}$)$^2$&	23.6\%	&23.2$\pm$1.2\%\\ \hline
\hbox{\vrule height 9pt depth 4.5pt width 0pt}
d,s&\mathstrut$-\frac{1}{2}$+ $\frac{2}{3}x$&	$-\frac{1}{2}$&	6($-\frac{1}{2}$  + $\frac{2}{3}$ x)$^2$ + 6\mathstrut$(\frac{1}{2})^2$&	30.3\%	&31.68$\pm$ 0.8\%\\ \hline
\hbox{\vrule height 9pt depth 4.5pt width 0pt}
b&\mathstrut$-\frac{1}{2}$+ $\frac{4}{3}x$&	$-\frac{1}{2}$&	3($-\frac{1}{2}$  + $\frac{2}{3}$x)$^2$ + 3($\frac{1}{2}$)$^2$&	15.3\%	&15.12$\pm$ 0.05\%\\ \hline
\end{tabular}
}
\label{t1}
\end{table}

\section{LEP I Results}\label{s7}
The final legacy papers describing and interpreting the results in the framework of the SM were published by the LEP EWWG in Physics Reports in 2006  for the \Z\  production at LEP I \cite{ALEPH:2005ab} and in 2013 for the W pair production at LEP II \cite{Schael:2013ita}.  Earlier results can be found  in  Refs.  \cite{Bardin:1999ak,Renton:2002wy,Altarelli:2004fq,Hollik:2006hd}, while later updates can be found in the reviews from the Particle Date Group \cite{Agashe:2014kda}
%
The four LEP experiments, shortly described in section \ref{s3}, collected between 1990 and 1995 a total of 17 million Z events distributed over seven CM energies with most of the luminosity taken at the peak. 
The total cross-section is given by $\sigma_{\rm tot}=(N_{\rm
sel}-N_{\rm bg})/(\epsilon_{\rm sel} \calL ) $, where $N_{\rm sel}$ is
number of selected events in a final state,   $N_{\rm bg}$ the number
of background events, $\epsilon_{\rm sel}$ the selection efficiency
including acceptance, and  $\cal L$ the integrated
luminosity.  We shortly discuss the uncertainties in these variables.
The combination of magnetic spectrometers with good tracking, electromagnetic and hadronic calorimeters and muon tracking allows a  good discrimination of $\qq$ from $\ell^+\ell^-$ final states
and a strong reduction of background, which was typically below 1\% for all final states (except for hadronic tau final states, where the background went up to 3\%).  
Since the background is largely independent of the LEP energy it provides a constant background, so it can be determined experimentally from off-peak measurements and is small, as discussed above. 

The luminosity is determined from small angle Bhabha scattering using the acceptance calculations and cross section from the
 the program BHLUMI, which was used by all experiments leading to a correlated common error from  the higher order uncertainties  in the Bhabha scattering cross section of  0.061\%  \cite{Ward:1998ht}.
 From calorimeters with high angular resolution silicon detectors the experiments obtained a lumino\-sity error of about 0.1\%, which led to an experimental error in the cross sections from the global fit  comparable to the theoretical uncertainty from the higher order corrections.

The acceptance  is limited largely by the geometrical acceptance. The electromagnetic calorimeters have typically a geometrical acceptance of $|\cos\theta| \le 0.7$, the muon trackers typically $|\cos\theta| \le 0.9$. For the hadronic final states the jets do not have a sharp angular edge for the acceptance, so the acceptance is limited by requi\-ring a fraction  of the total CM energy to be visible in the detector (typically 10\%). Since the simulation programs of  the \Z\  decays and the detector simulation\footnote{Details about the simulation software  can be found in Ref. \cite{ALEPH:2005ab}.} are   realistic inside the acceptance, the total efficiency can be extrapolated reliably to the full acceptance. 
Inside the acceptance the trigger efficiency is usually high, since events can be triggered by a multitude of signals, like track triggers, calorimetric triggers and combinations thereof.
The selection efficiencies inside the acceptances are high, above 95\% for electrons and muon pairs and 70 - 90\% for tau pair final states.
\begin{figure}
\begin{center}
\includegraphics[width=0.4242\textwidth,height=0.22\textheight]{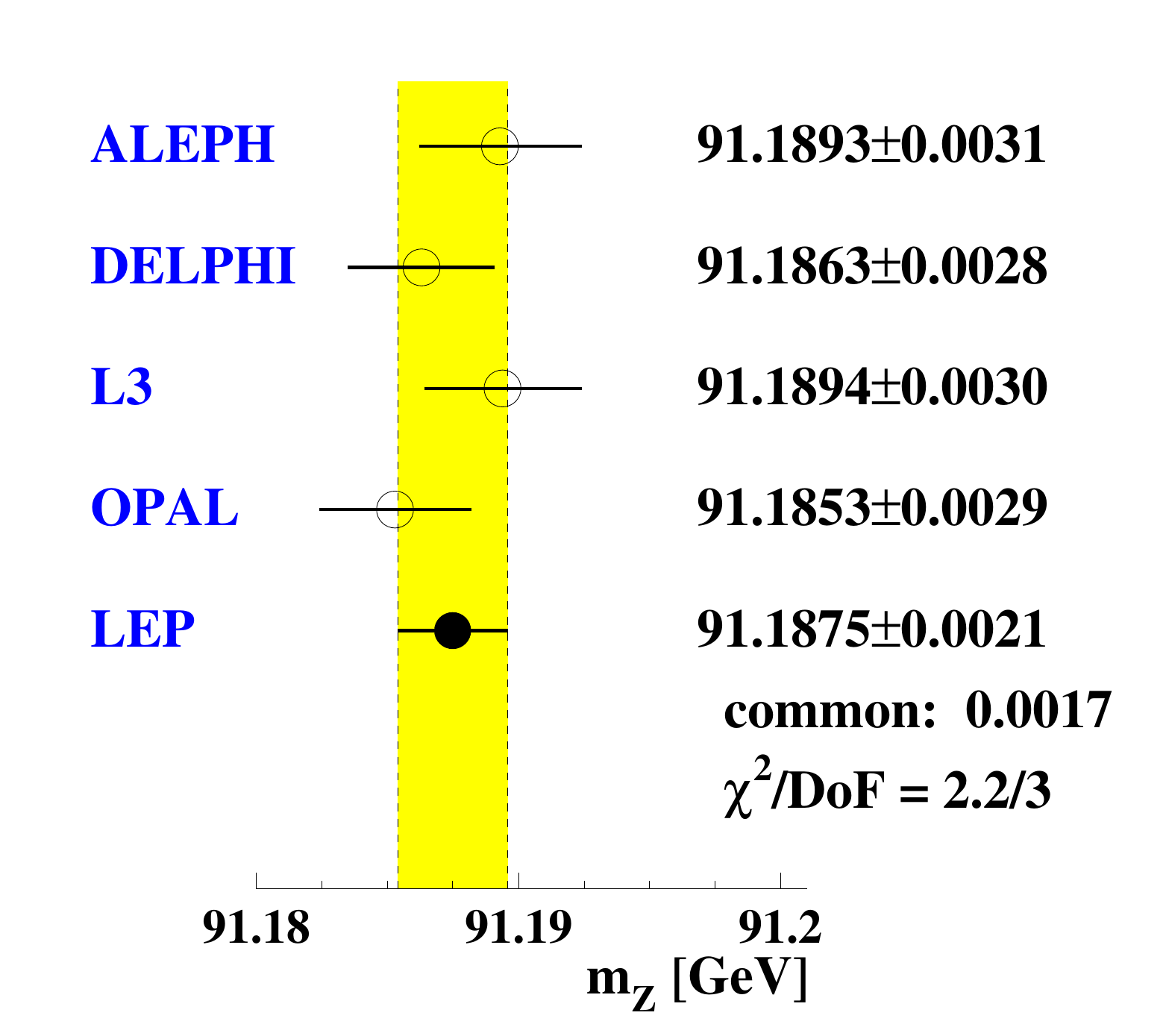}\hspace*{10mm}
\includegraphics[width=0.4242\textwidth,height=0.22\textheight]{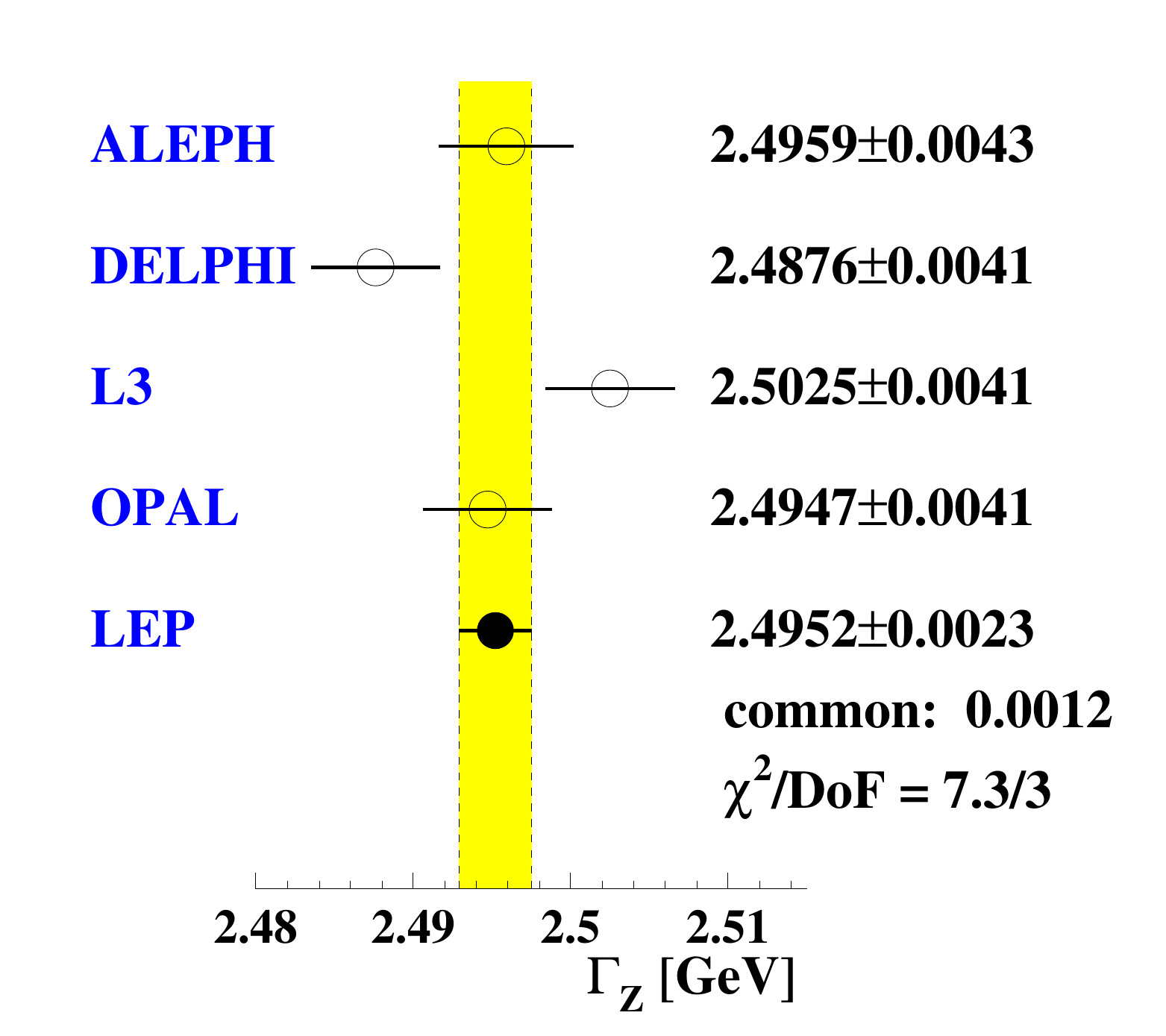}
\includegraphics[width=0.4242\textwidth,height=0.22\textheight]{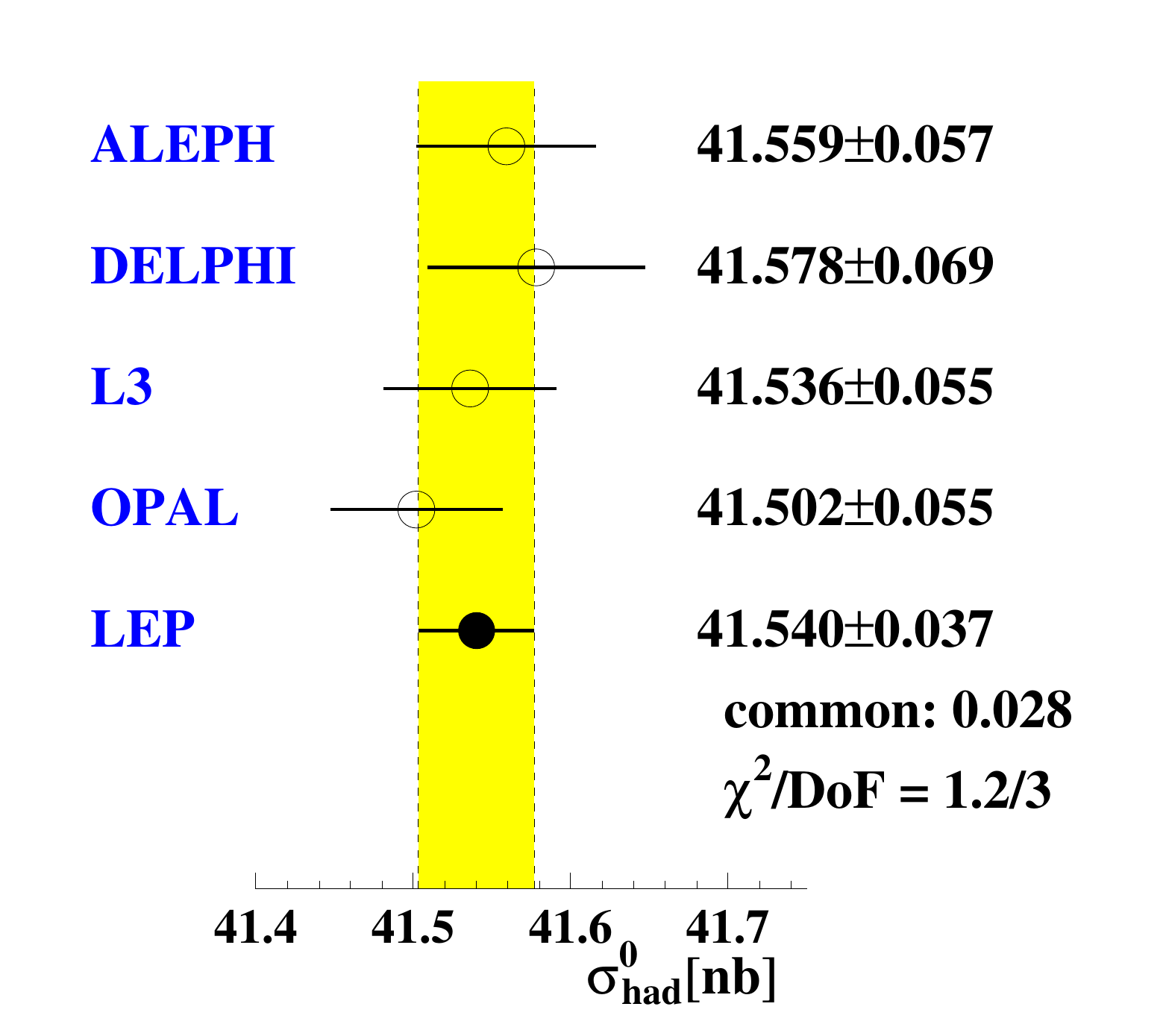}\hspace*{13mm}
\includegraphics[width=0.4242\textwidth,height=0.22\textheight]{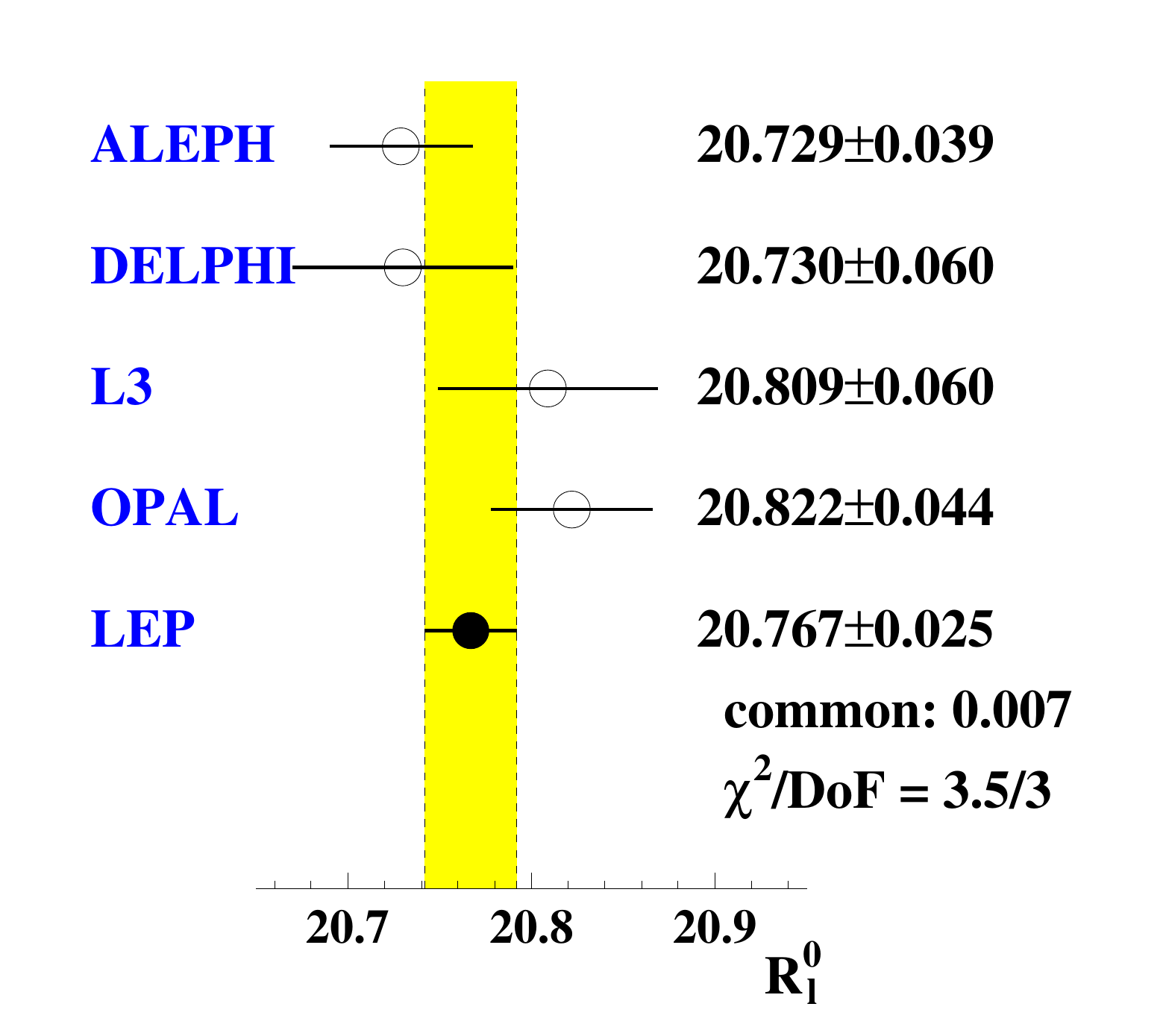}\vspace*{-2mm}
\caption[]{ The fitted values of the mass, width (top row), peak cross section and ratio of hadronic to leptonic width (bottom row) of the \Z\ boson. From Ref \cite{ALEPH:2005ab}.}
\label{f4}
\end{center}
\end{figure}
   The symmetric Breit-Wigner function can be des\-cribed by the mass, the width and the peak height. The leptonic cross sections can be parametrized by the ratio of hadronic and leptonic widths: $\Rl=\Gqq/ \Gll$. Since lepton universality was compatible with all observations, we quote only results including lepton universality. 
The fitted values for these parameters from the various experiments and their combination are shown in Fig. \ref{f4}. 
One observes that for the combined values of the experiments the common systematic errors are large in case of the hadronic final states, but for the leptonic final state the statistical error is still significant. The systematic errors on mass and width are dominated by the uncertainty of  the LEP energy (around 2 MeV, as discussed in Section \ref{s2}) and for the cross section by the luminosity error  discussed above.

The combined fit to all data requires a knowledge of the correlated errors between the observables and the experiments. These correlations can be taken into account by minimizing $\chi^2=\Delta^T V^{-1} \Delta$, where $\Delta$ is the vector  containing the N residuals between the N measured and fitted values, $V$ is the NxN error matrix  where the diagonal elements $\sigma_{ii}^2/O_i^2$ represent   the relative total error squared for observable $O_i$ and the off-diagonal elements $\sigma_{ij}^2/(O_i Q_j)$ the relative correlated error. E.g. for the correlated error of the Bhabha luminosity of 0.061\% is added in quadrature to all off-diagonal elements of observables depending on the luminosity. This method was pioneered for  \ee\ annihilation data from the DORIS and PETRA colliders at DESY and the TRISTAN collider at KEK, where the tail of the  \Z\ resonance increases the hadronic cross section already by 50\% at the highest energy of 57 GeV \cite{D'Agostini:1989cz}. The complete correlation matrices for all LEP data can be found in the final report from the EWWG \cite{ALEPH:2005ab}.

\section{Constraints on the SM}\label{s8}
\begin{figure}
  \begin{center}
\begin{minipage}{0.49\textwidth}
\includegraphics[width=\textwidth,height=1.02\textwidth]{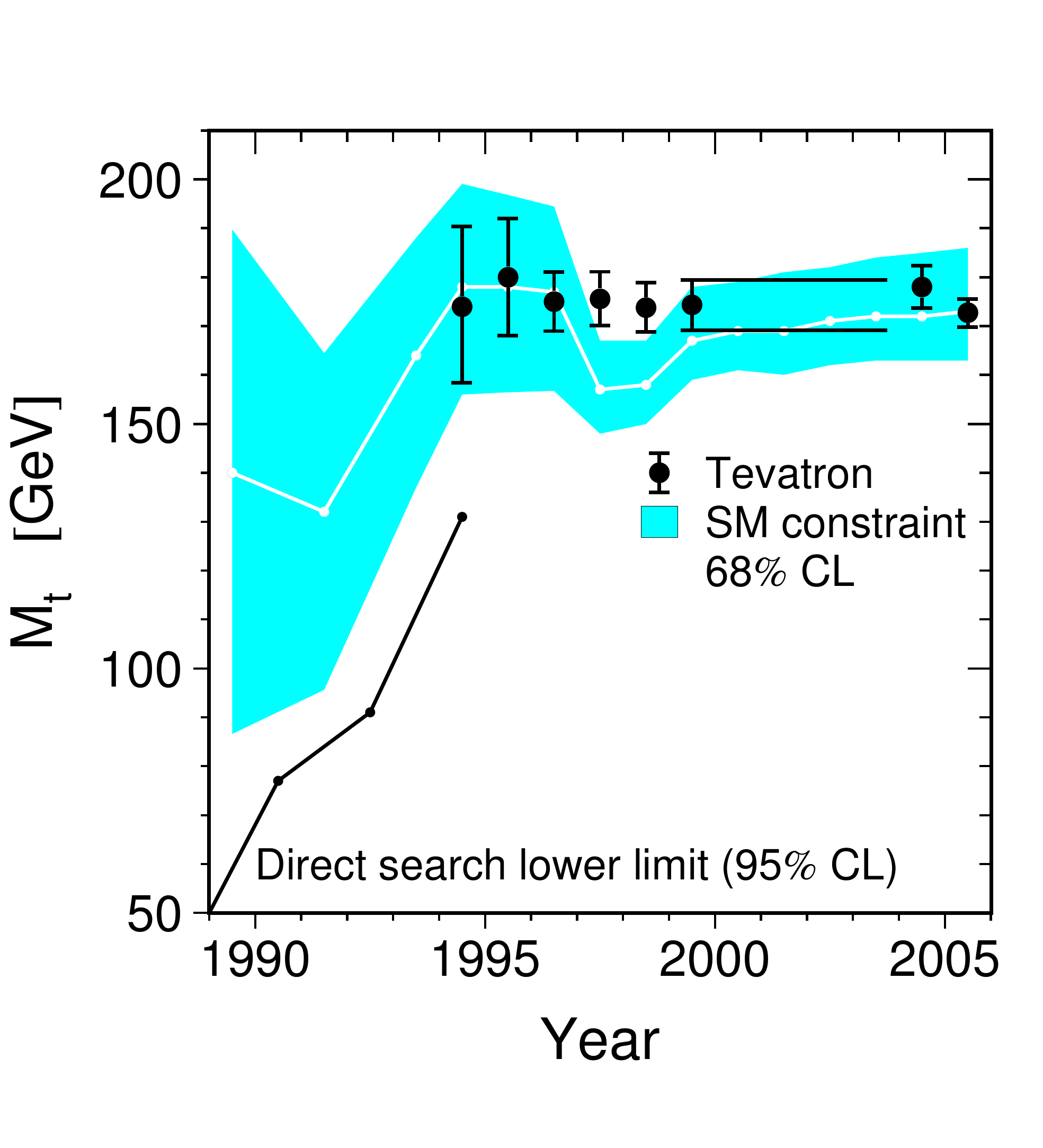}
\end{minipage}\hspace*{4mm}
\begin{minipage}{0.48\textwidth} 
\includegraphics[width=\textwidth,height=0.87\textwidth]{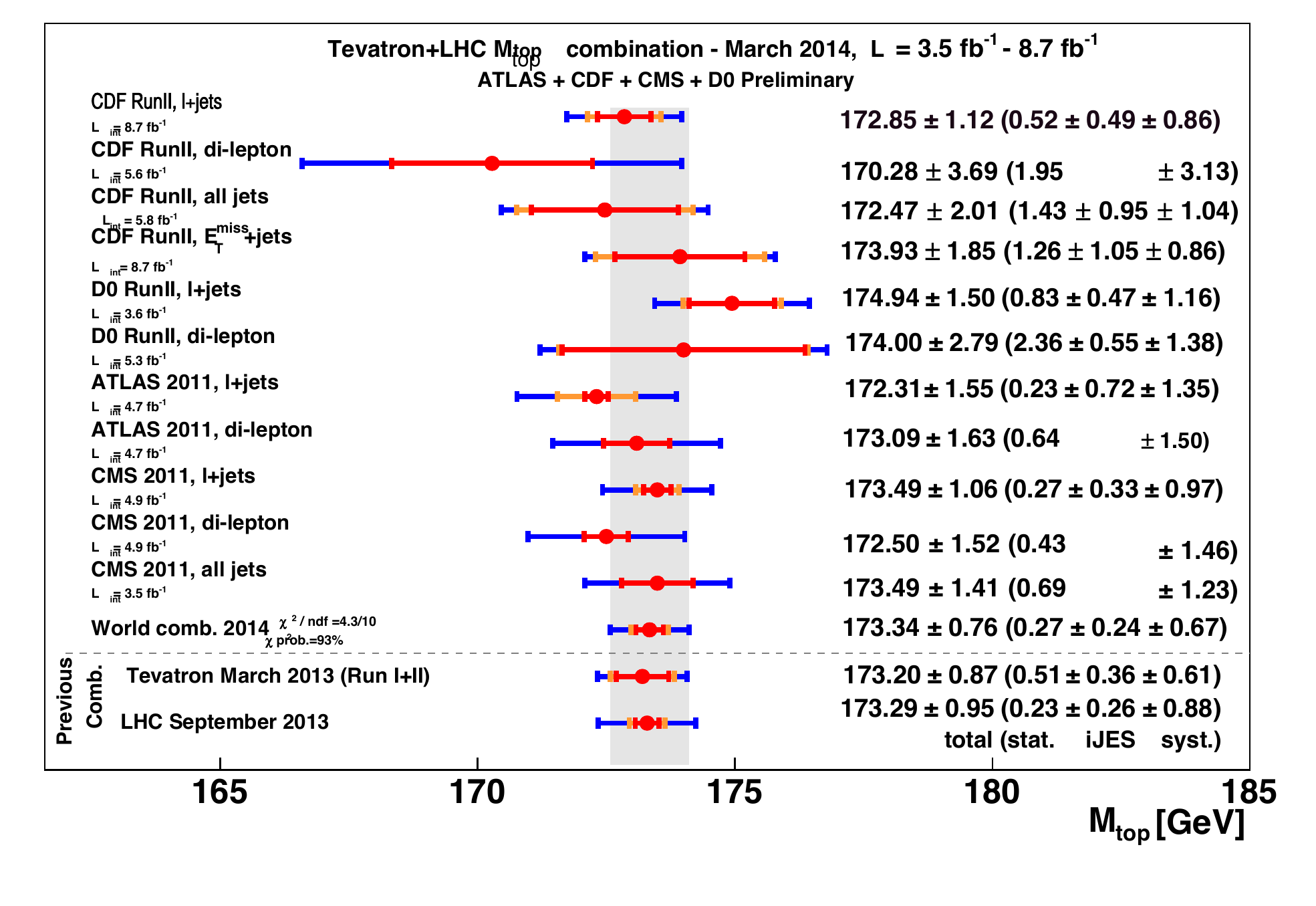}
\end{minipage}\vspace*{-5mm}
\center{\hspace*{15mm}(a)\hspace*{0.48\textwidth}(b)}
    \caption[]{(a): The measured top quark mass as function of time \cite{ALEPH:2005ab}. The indirect determinations from the electroweak fits (shaded area)  to LEP data predicted a heavy top quark mass  before it was discovered at the Tevatron (data points).  (b): A summary of direct top quark measurements \cite{ATLAS:2014wva}. } \label{f5}
  \end{center}
\end{figure}
The measurements of the cross sections and asymmetries discussed above can all be predicted in the SM, if one knows the three gauge couplings, the gauge boson masses and the masses of the top quark and Higgs boson. Since the electromagnetic and weak couplings are related via the gauge boson masses, only two coupling constants are needed: $\alpha(\MZ)$ and  $\alpha_s(\MZ)$. Furthermore, $\MW$ can be traded for  $\GF$, which 
was recently measured from the muon lifetime to 0.5 ppm:   \mbox{$\GF = 1.1663787(6) \cdot 10^{-5}\, {\rm GeV}^{-2}$} \cite{Tishchenko:2012ie}.
 This value is precise enough to be considered a constant in the fit. The masses of the light fermions have only a small effect on the cross section and their effect can be calculated with sufficient precision.   $\alpha (\MZ)$ is in principle known from the running from its low energy value, but the loop corrections including quarks have a significant uncertainty. Therefore, the hadronic contribution for 5 quarks to $\dalhad$ is taken as a parameter in the fit (instead of $\alpha (\MZ)$)  with the constraint from the experimental knowledge on $\dalhad$. The SM parameters to be fitted to the measured observables are then: $\MZ$, $\mt$, $\MH$, $\alpha_s$, $\dalhad$. 
 Given  these parameters all  observables can be calculated, e.g. with the programs TOPAZ0~\cite{Montagna:1998kp},  ZFITTER~\cite{Arbuzov:2005ma} or GAPP \cite{Erler:1999ug}.
\begin{figure}
\begin{center}
\begin{minipage}{0.46\textwidth}
\includegraphics[width=\textwidth,height=0.46\textheight]{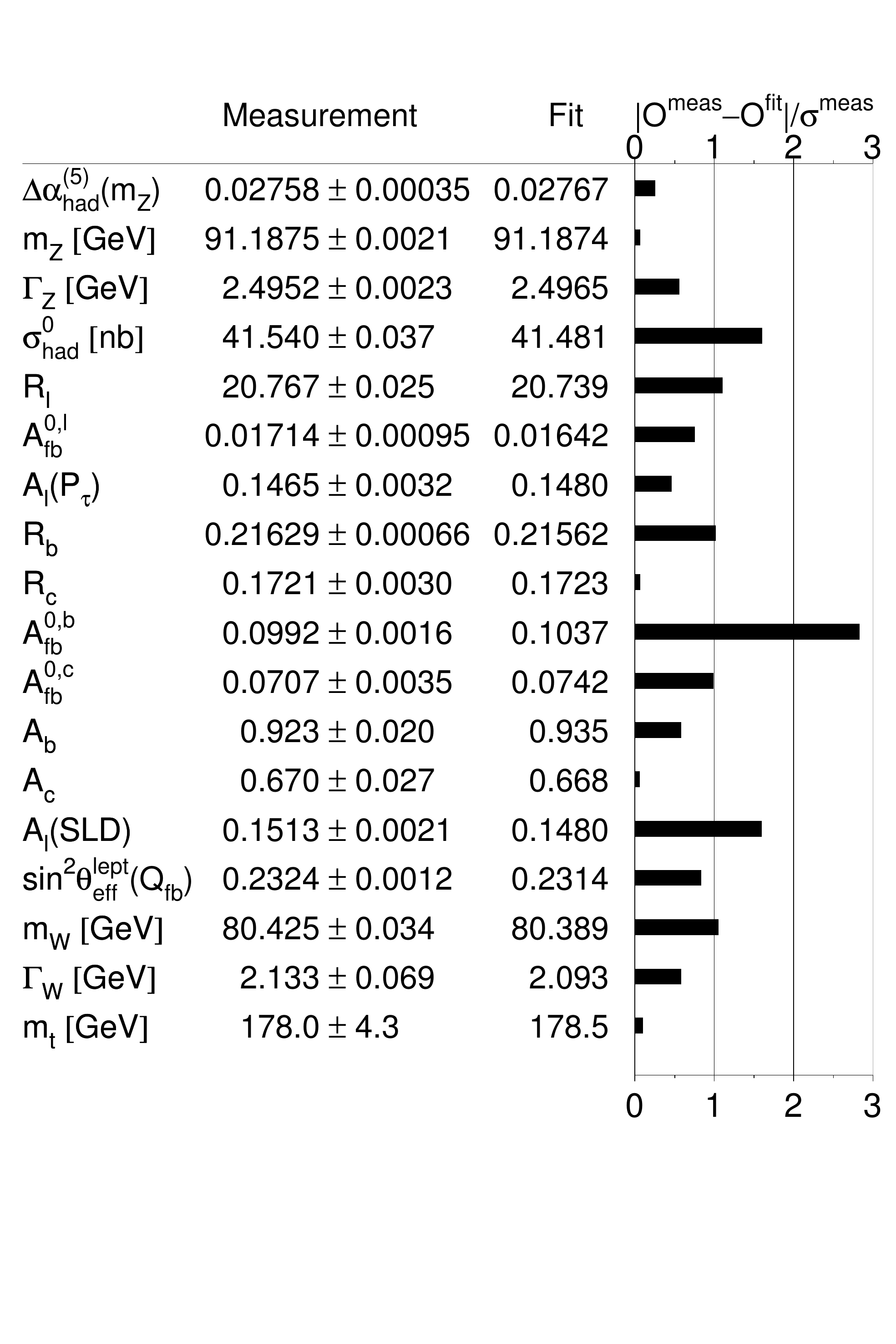}\vspace*{-15mm}
\end{minipage}\hspace*{10mm}
\begin{minipage}{0.46\textwidth}
\includegraphics[width=\textwidth,height=0.37\textheight]{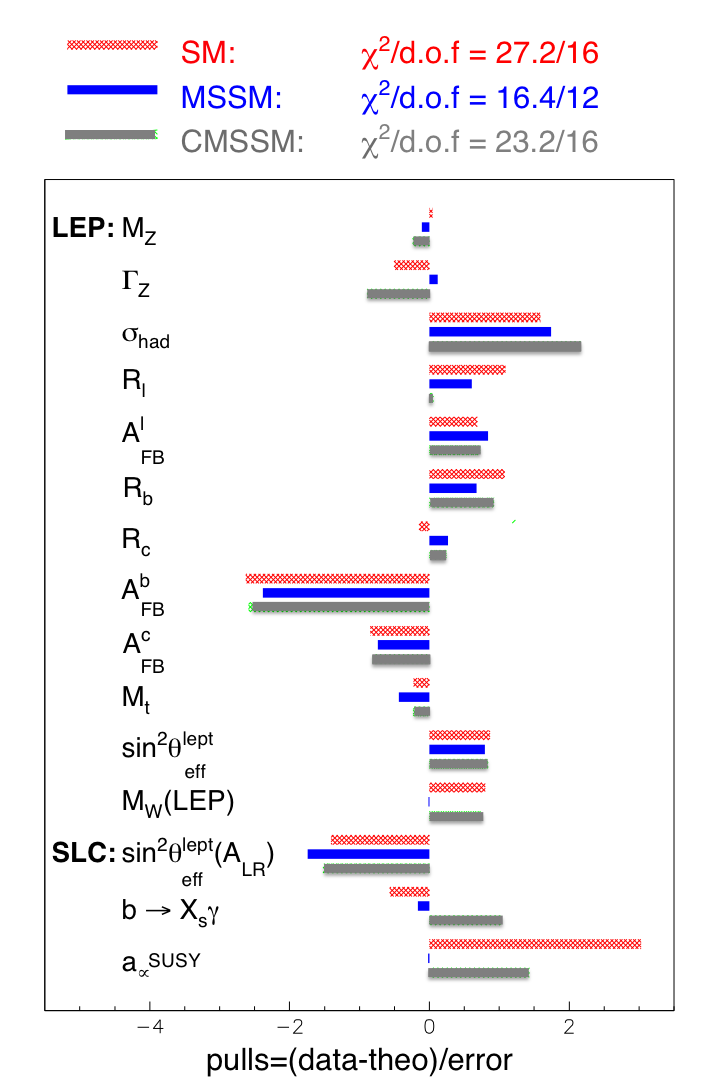}
\end{minipage}
\center{\hspace*{5mm}(a)\hspace*{0.5\textwidth}(b)}
\caption[]{(a): A comparison of the measured and calculated values of the precision electroweak observables and a graphical presentation of the difference, expressed in standard deviations (``pulls''). The fit has a $\chi^2/d.o.f$ of 18.3/13, corresponding to a probability of 15\%.
(b): A comparison of the pulls in the SM, the minimal supersymmetric SM (MSSM) and constrained MSSM (CMSSM). }
\label{f6}
\end{center}
\end{figure}
%
The quadratic top quark dependence of the loop corrections to the gauge boson masses led quickly to first estimates of the top mass from the precise Z boson mass measurements, as shown in Fig. \ref{f5}a. These top mass estimates were confirmed later by  direct measurements, as shown by the data points from the Tevatron experiments in Fig. \ref{f5}a, which in turn agree with the LHC measurements, as shown in Fig. \ref{f5}b. 

From a fit to the Z-pole data and preliminary data for $\mt$ and $\MW$  the EWWG finds for these parameters \cite{ALEPH:2005ab}: $\MZ=91.1874\pm 0.0021$, $\mt=178.5\pm 3.9$ GeV, $\MH=129^{+74}_{-49}$ GeV, $\alpha_s=0.1188\pm 0.0027$ and $\dalhad=0.02767\pm 0.00034$\footnote{With newer data  the value quoted in the Particle Data Book \cite{Agashe:2014kda} has a considerably smaller error: $0.02771 \pm 0.00011$.}. These five parameters describe the data quite well, as shown 
 in Fig. \ref{f6}a, which displays the difference between the calculated and observed values of the observables.  
\begin{figure}
  \begin{center}
\vspace*{-1mm}
\begin{minipage}{0.49\textwidth}
\includegraphics[width=\textwidth,height=0.85\textwidth]{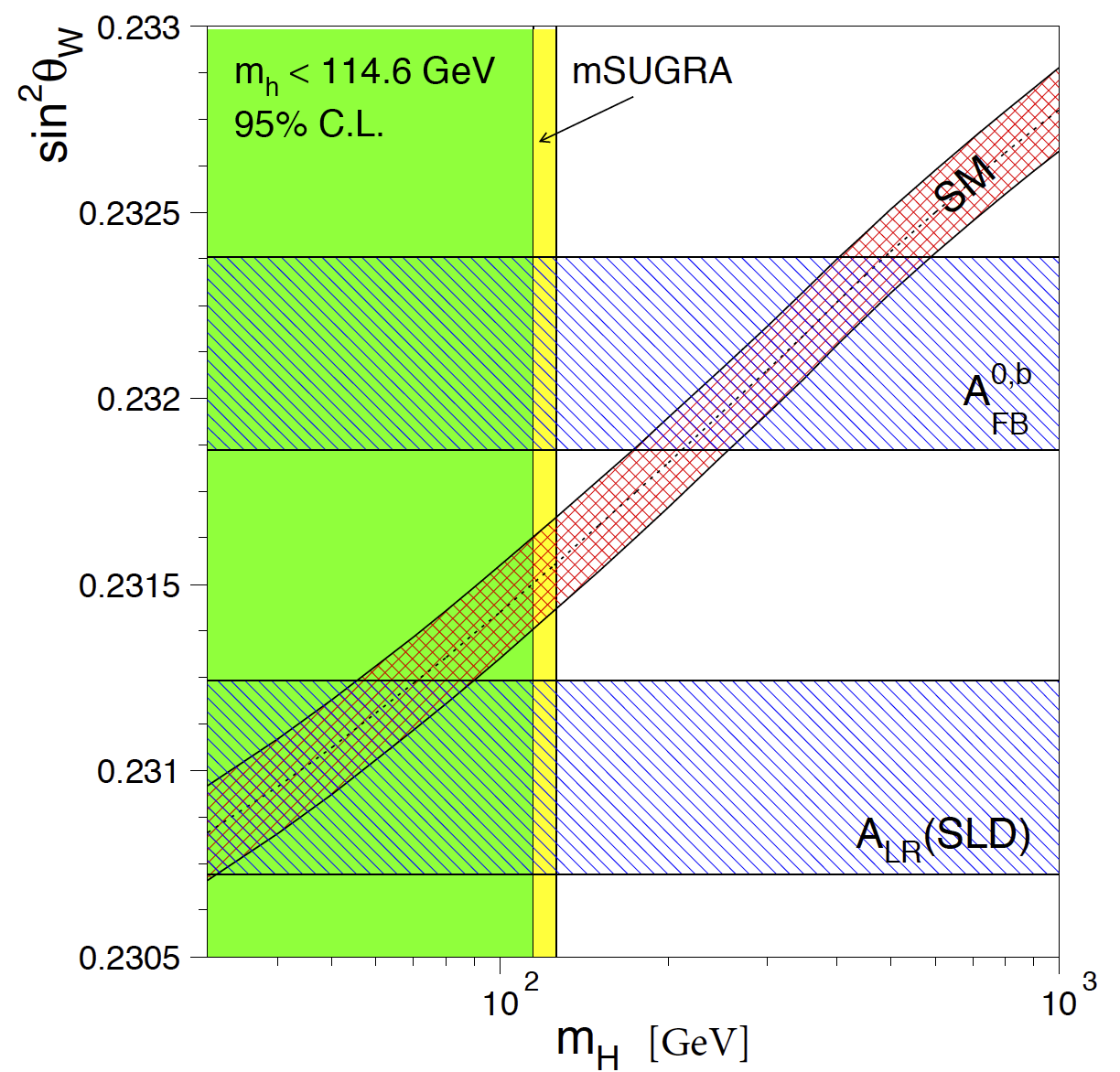}
\end{minipage}\hspace*{4mm}
\begin{minipage}{0.48\textwidth} 
\includegraphics[width=\textwidth,height=0.99\textwidth]{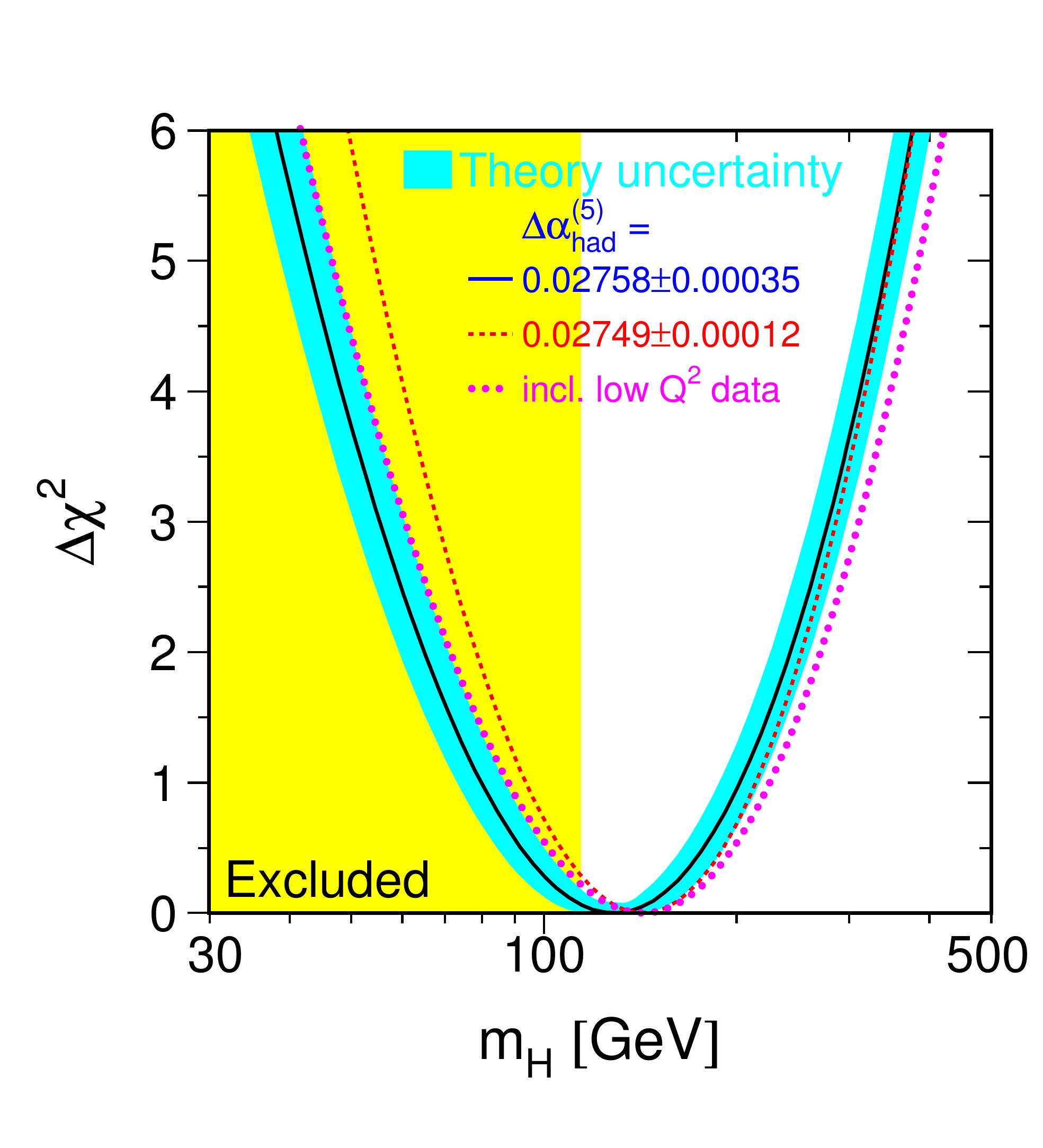}
\end{minipage}\vspace*{-5mm}
\center{\hspace*{20mm}(a)\hspace*{0.4\textwidth}(b)}
    \caption[]{(a): The values of $\sinw$ versus the Higgs mass. The two horizontal bands correspond to the \sinw\ values from  $\ALR$ and $\Afbzb$. The diagonal band corresponds to the SM prediction for the parameters from the global fit. The shaded (green) area for $\MH<114.3$ GeV is excluded by LEP data. (b): The $\Delta\chi^2$ distribution as function of the Higgs mass from the LEP I and SLC data before the Higgs boson discovery, but including the constraints from $\MW$ and $\mt$ \cite{ALEPH:2005ab}. The minimum corresponds to $M_h=129\pm^{74}_{49}$ GeV.} \label{f7}
  \end{center}
\end{figure}
The largest pull of 2.8$\sigma$ is caused by the forward-backward asymmetry of the b quarks, followed by 1.6$\sigma$ for the peak cross section and the left-right polarization asymmetry from SLC.  
%
The correlation between the Higgs mass and \sinw is demonstrated  in Fig. \ref{f7}a, where the diagonal shows the SM prediction. The two horizontal bands show  the $\sinw$ values from $\ALR$ and $\Afbzb$, which lead to quite different Higgs mass values, as is apparent from the crossing with the SM prediction.
 The narrow vertical (yellow) band shows the expected Higgs mass in the supersymmetric extension of the SM.
The Higgs boson mass observed at the LHC falls inside this SUSY band, which crosses the SM prediction at a $\sinw$ value close to the value from the averaged asymmetry. The indirectly measured Higgs boson mass falls also inside this band,
as demonstrated by the ``blue-band'' plot in Fig. \ref{f7}b, although the errors are large in this case: $M_h=129\pm^{74}_{49}$ GeV.
In addition to the discrepancy in the asymmetries 
the anomalous magnetic moment of the muon  $a_\mu$ shows a 3$\sigma$ deviation from the SM \cite{Jegerlehner:2009ry}.  Supersymmetric loop corrections to  $a_\mu$ reduce the observed difference between theory and experiment, so many groups have tried to improve the fit in supersymmetric extensions of the SM (see Refs. \cite{Heinemeyer:2004gx,Hollik:2006hd} for reviews), both in the Minimal Supersymmetric SM (MSSM) and in the constrained version (CMSSM)\footnote{With the present lower limits on   SUSY masses no improvement of $a_\mu$ is possible in the CMSSM,  for details, see e.g. Ref. \cite{Beskidt:2012sk} and references therein. }. Here minimal means the minimum extension of the SM, i.e. one superpartner for each SM particle and a minimal Higgs sector of two Higgs doublets. In the CMSSM one assumes in addition unification of gauge couplings and SUSY masses at the GUT scale. 
Unfortunately,  the largest pull from  $\Afbzb$ does not improve with SUSY, as shown by the ``pulls'' in Fig. \ref{f6}b \cite{deBoer:2003xm}. Although the $\chi^2$ is smaller in the (C)MSSM, the probability stays similar, because of the larger number of parameters. 
\subsection{Constraints on the SM after the Higgs discovery}\label{s8a}
The global fits have been repeated after the Higgs discovery  and the results have been described by Erler and Freitas in the electroweak review of the Particle Data Group \cite{Agashe:2014kda}. 
Also newer values from $\MW$, $\GF$ and $\mt$ have been included. The anomalous muon magnetic moment has been fitted as well. 
The global fit including the measured top  quark and Higgs boson masses yields a good $\chi^2/d.o.f$ of 48.3/44. The probability to obtain a larger $\chi^2$ is 30\%. 

\begin{figure}[t]
\begin{center}
\includegraphics[width=0.48\textwidth]{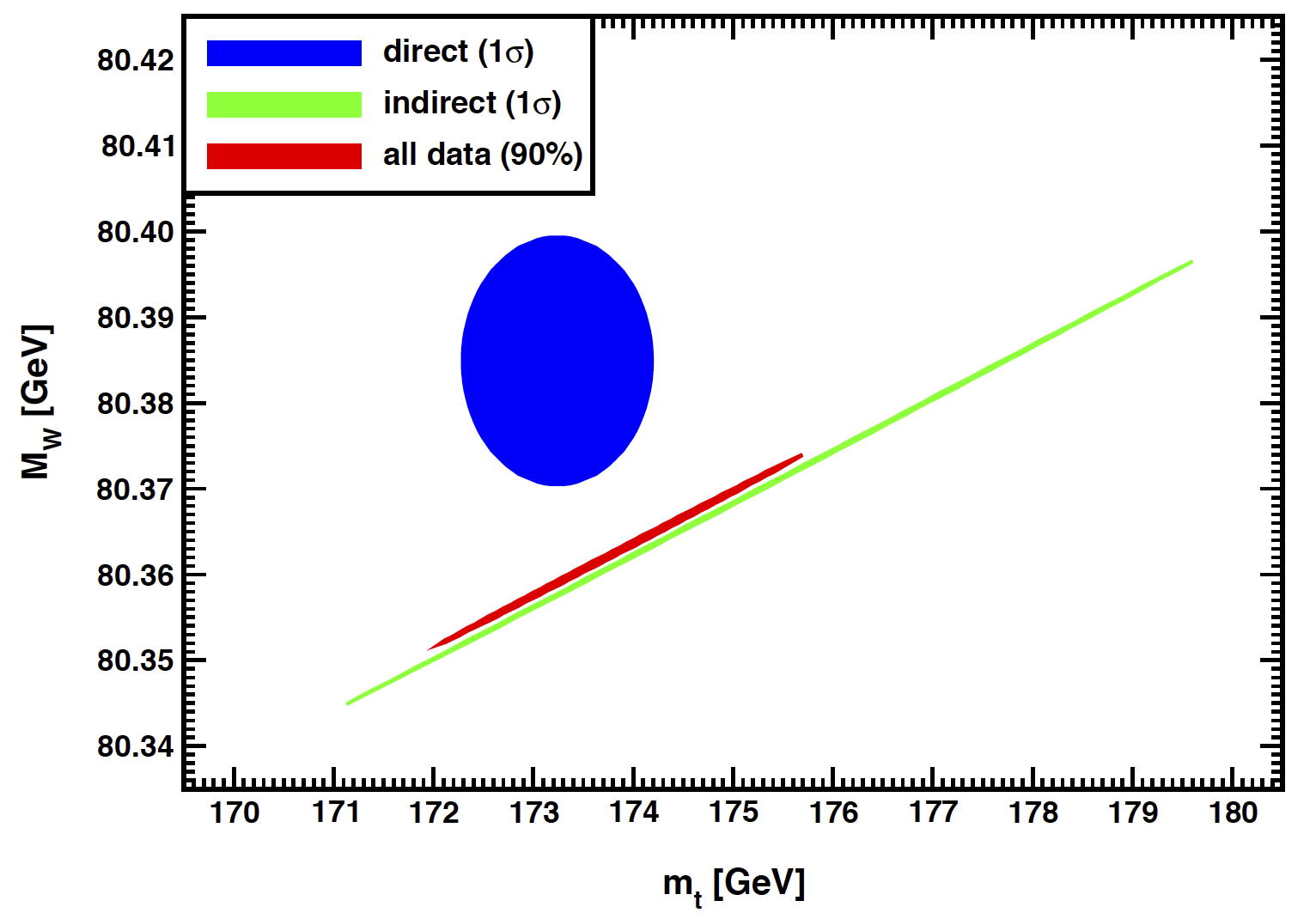}\hspace*{5mm}
\includegraphics[width=0.48\textwidth]{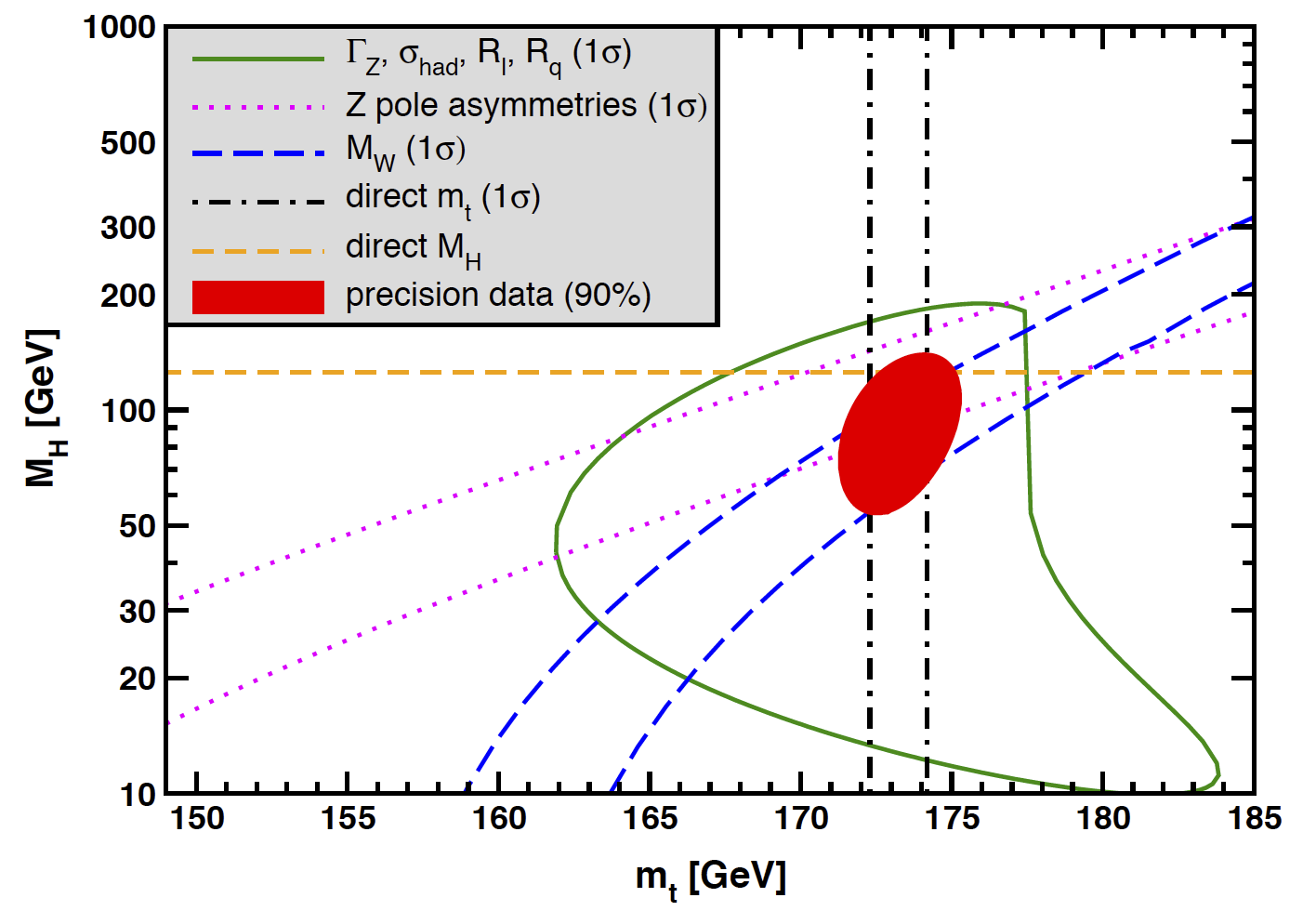}\vspace*{-2mm}
\center{\hspace*{10mm}(a)\hspace*{0.5\textwidth}(b)}
\caption{(a):  Allowed 1$\sigma$ contours with a probability of 39.35\% in the  $\MW$ versus  $\mt$ plane 
for the direct (dark (blue) ellipse)  and indirect measurements (light (green)``line''). The dark (red) ``line'' is the 90\% C.L. contour ($\Delta \chi^2$ = 4.605) allowed by all data. From Ref.  \cite{Agashe:2014kda}.
(b): Allowed 1$\sigma$ contours with a probability of 39.35\% in the  $\MH$  versus  $\mt$ plane for various observables.  The  dark (red) ellipse corresponds to the 90\% C.L. contour ($\Delta \chi^2$ = 4.605) from a global fit to all data. From Ref.  \cite{Agashe:2014kda}.  }
\label{f8}
\end{center}
\end{figure}

To check the consistency between direct mass measurements of $\MW$, $\mt$ and $\MH$ and the SM predictions via indirect measurements we show two examples from the Particle Data Group \cite{Agashe:2014kda}. Fig. \ref{f8}a shows the SM prediction for \MW\ versus the top quark mass as the light (green) diagonal contour, which shows the quadratic dependence of the gauge boson mass on the top quark for a Higgs boson of 125 GeV. This contour almost collapses into a line, because the precisely measured Higgs mass was included in the fit. Other\-wise the line would have been a band in this plane, since higher Higgs boson masses would shift this line parallel to lower W masses.   The direct measurements of  $\MW$ and $\mt$  are bounded by the dark  (blue) ellipse. These contours of the direct and indirect measurements (green and blue) correspond to 1$\sigma$  with   a probability of 39\%. 
Combining the direct and indirect measurements leads to the dark (red) ``line'',  
for which $\Delta\chi^2=4.61$ or a probability of 90\% was chosen. 
The value of the directly measured  \MW\ mass is 1.5$\sigma$ above the SM prediction \cite{Agashe:2014kda}, which implies some tension between \MW\ and \MH, since lower values of \MH would shift the SM prediction upwards. This tension is also visible in
  Fig. \ref{f8}b, which shows  the allowed 1$\sigma$ contours in the  $\MH$ versus  $\mt$ plane from various indirect measurements. The direct measurements are indicated by the horizontal and vertical lines.  The error on the Higgs mass is   not visible on this scale. The  dark (red) ellipse corresponds to the 90\% C.L. contour ($\Delta \chi^2$ = 4.605) from a global fit to all data  \cite{Agashe:2014kda}.  The central value of the ellipse  (indirect measurements) is slightly below  the direct measurement of the Higgs boson mass, since the slightly high value of \MW\  pulls the Higgs mass to lower values. Although the indirectly measured Higgs mass is not precise, it indicated for the first time that a Higgs boson is needed with a mass around the electroweak scale, a value  predicted by SUSY \cite{Djouadi:2005gj}. In the SM the Higgs boson mass is not predicted \cite{Djouadi:2005gi}.

\section{LEP II results}\label{s9}
The LEP II data allowed to investigate the selfcoupling of the gauge bosons by studying  W pairs, which can be produced in \ee\ annihilation via t-channel neutrino-exchange and s-channel photon, Z and Higgs exchange. As mentioned in Sect. \ref{s5} the Higgs exchange is needed to compensate the divergences from the longitudinal components of the gauge bosons. One can indeed verify by explicit calculations that  the amplitudes cancel at high energies, i.e. $A_\nu +A_\gamma+A_Z$=-$A_H$.
\begin{figure}[t]
\begin{center}
\begin{minipage}{0.48\textwidth}
\includegraphics[width=\textwidth]{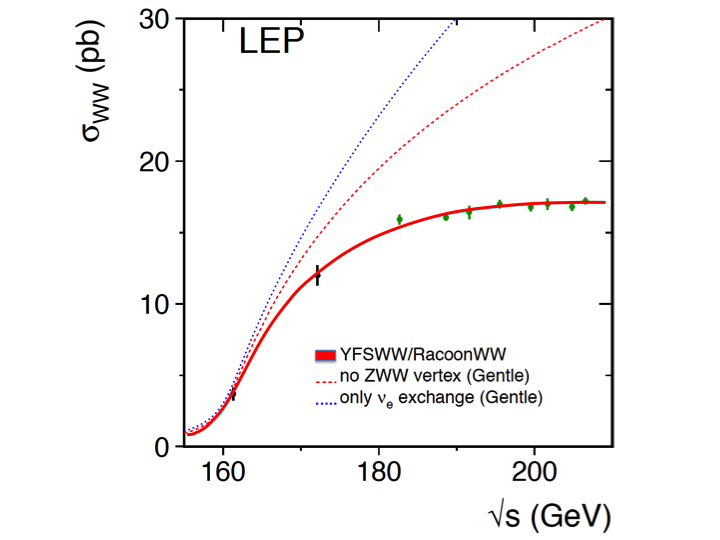}
\end{minipage}\hspace*{5mm}
\begin{minipage}{0.48\textwidth}
\includegraphics[width=\textwidth]{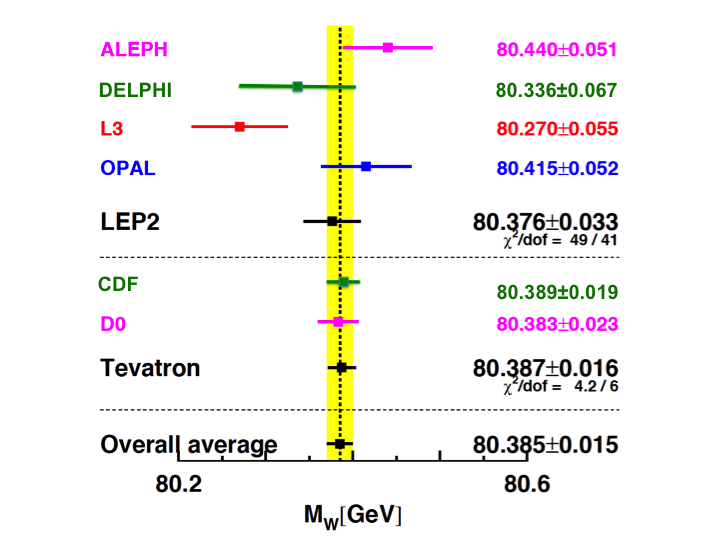}
\end{minipage}
\center{\hspace*{15mm}(a)\hspace*{0.48\textwidth}(b)}
\caption{(a): The W pair production cross section at LEP II as function of the centre-of-mass energy. Without ZWW vertex the cross section would diverge as function of energy, as shown by the dotted lines  for the cases that only the t-channel neutrino exchange or both, neutrino and photon exchange, (``no ZWW'') would be present. (b) A comparison of the directly measured W boson masses. From Ref.  \cite{Schael:2013ita}.} 
\label{f9}
\end{center}
\end{figure}
However, the Higgs exchange is proportional to $m_e\sqrt{s}/M_W^2$, so this term becomes only important for $\sqrt{s}\approx \MW^2/m_e \approx 10^7$ GeV. At LEP II energies the longitudinal cross section can be neglected and only   $A_\gamma$, $A_\nu$ and $A_Z$ are important.  
Each of them increases with the energy squared, but  $A_Z$  interferes negatively with the other amplitudes. The energy dependence of $A_\nu$,
$A_\nu + A_\gamma$ and the total cross section are displayed in Fig. \ref{f9}a. 
The negative interference leads to a   rather  slow  energy dependence of the total W pair production cross section by virtue of the fact that the triple gauge boson vertex in $A_Z$ has the same gauge coupling as  the coupling to fermions, a feature imposed by the gauge invariance of the SM.  
 One observes excellent agreement between the SM prediction and data. The shape of the cross section in Fig. \ref{f9}a is sensitive to $\MW$. Combining this  shape with invariant mass distributions of W final states leads to: $\MW= 80.376 \pm 0.033$ GeV and $\GW = 2.195 \pm 0.083$ GeV \cite{Schael:2013ita},  which agrees with mass measurements at the Tevatron, as shown in Fig. \ref{f9}b. The  world average of the directly measured W masses ($\MW=80.385\pm 0.015$ GeV)  is slightly higher than the indirectly measured W masses from the global electroweak fit ($\MW=80.363\pm 0.006$ GeV), as shown before in Fig. \ref{f8}a, but the discrepancy is only at the 1.5$\sigma$ level, as discussed before.

\section{QCD Results}\label{s10a}
The LEP I data were an eldorado for studying QCD given the  high Z boson  cross section and large branching ratio  into hadrons ($\approx$70\%, see Table \ref{t1}).  Among the milestones: i) a direct demonstration of the self interaction of gluons, thus confirming experimentally the basis for asymptotic freedom; ii) The precise experimental measurement of the strong coupling constant; iii) From a comparison with lower energy data evidence for the running of the bottom quark mass and the running of the strong coupling constant. We shortly describe these impressive results. 
\begin{figure}
\begin{center}
\includegraphics[width=0.46\textwidth,height=0.45\textwidth]{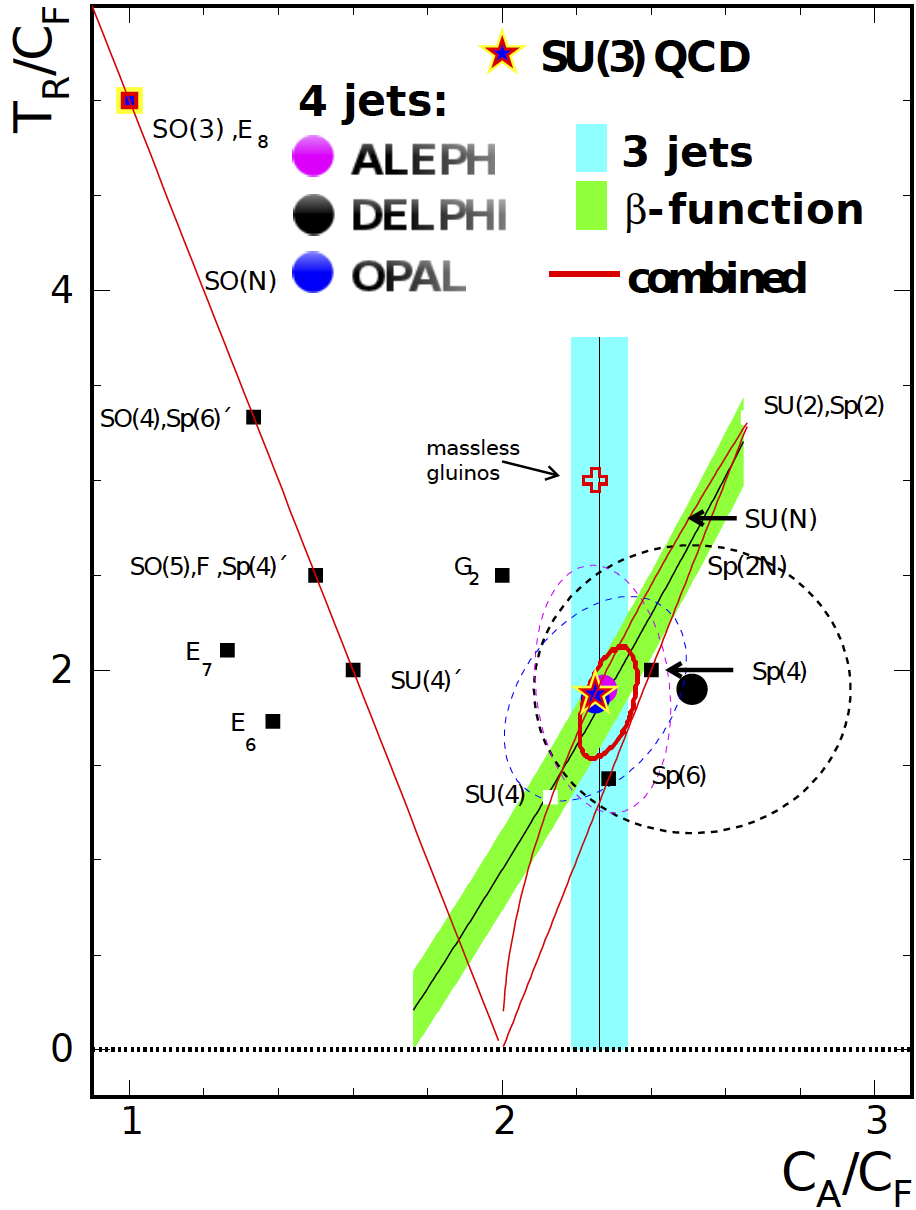}\hspace*{6mm}
\includegraphics[width=0.50\textwidth,height=0.47\textwidth]{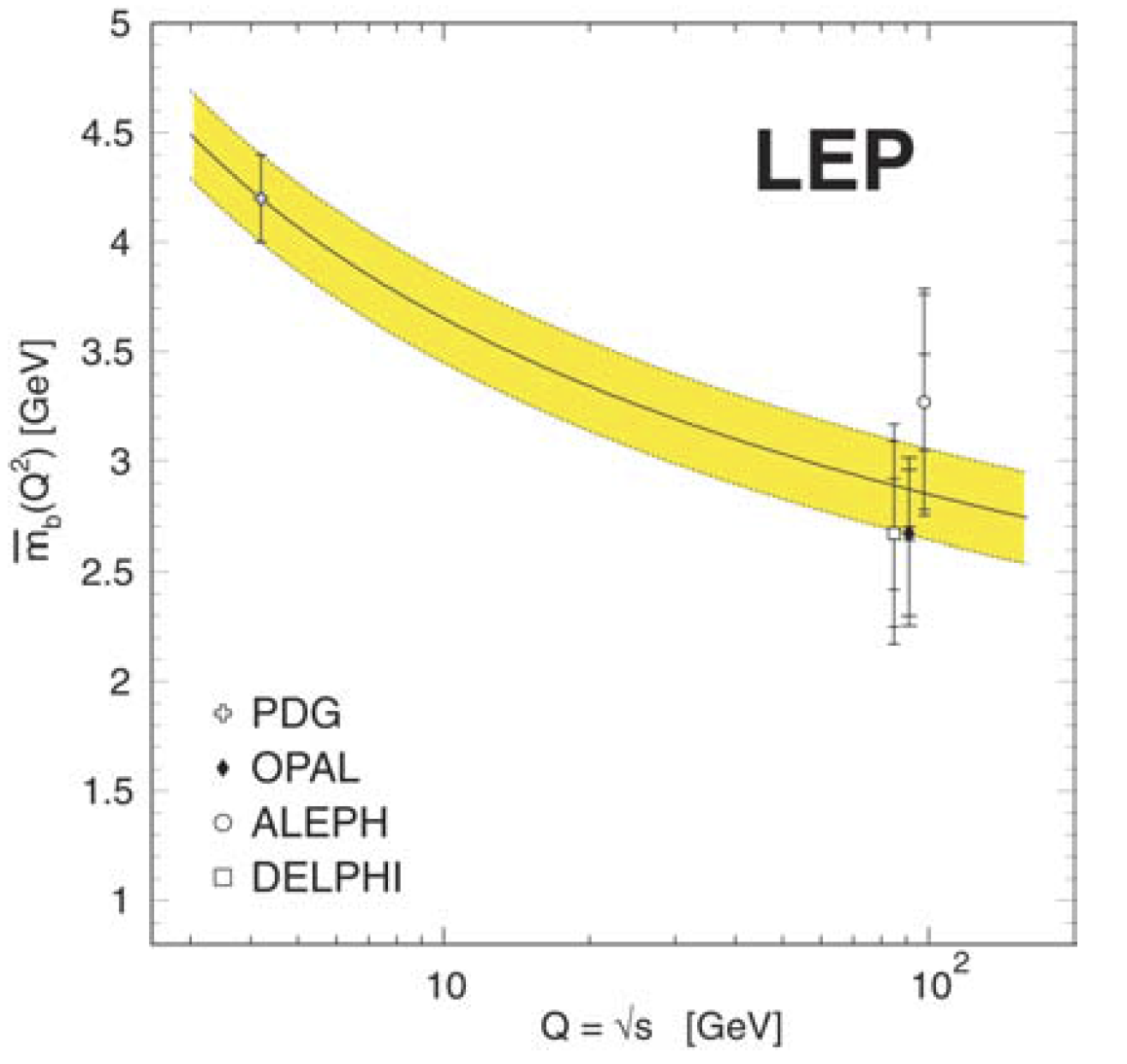}
\center{\hspace*{10mm}(a)\hspace*{0.44\textwidth}(b)}
\caption{(a):  $T_R/C_F$ versus $C_A/C_F$, where the colour factors $T_R$, $C_F$ and $C_A$ are associated with $g \rightarrow q \overline{q}$, $q \rightarrow q g$ and  $g \rightarrow g g$, respectively.  The combined fit to all data ((dark (red) ellipse) agrees with the SU(3) group from QCD, but excludes many other groups, see Ref. \cite{Abdallah:2005cy} for details and further references. (b): The running of the b quark mass. From Ref. \cite{Zerwas:2004ng}.}
\label{f10}
\end{center}
\end{figure}

\subsection{The gluon self interaction}\label{s10b}
Four jet events in \ee\ annihilation originate either from a radiation of two gluons or radiation of a single gluon with subsequent spitting either  into two quarks or two gluons. All three contributions have a different angular distribution and different cross section, so with the clean and high statistics of 4-jet events at LEP one can disentangle the various contributions. The contribution from the triple gluon vertex is clearly established \cite{Adeva:1990nw,Abreu:1990ce,Abreu:1993vk,Decamp:1992ip} and agrees with the SU(3) prediction,  as shown in Fig. \ref{f10}a by the filled circles.  In addition, the gluon self-coupling increases the gluon jet multiplicity and changes the averaged thrust with increasing energy, as determined by the beta function  of the RGE. Combining all these measurements \cite{Abdallah:2005cy}  constrains the gauge group of the strong interactions to SU(3),  as shown in Fig. \ref{f10}a.
\begin{figure}
  \begin{center}
    \includegraphics[width=0.48\textwidth,height=0.445\textwidth]{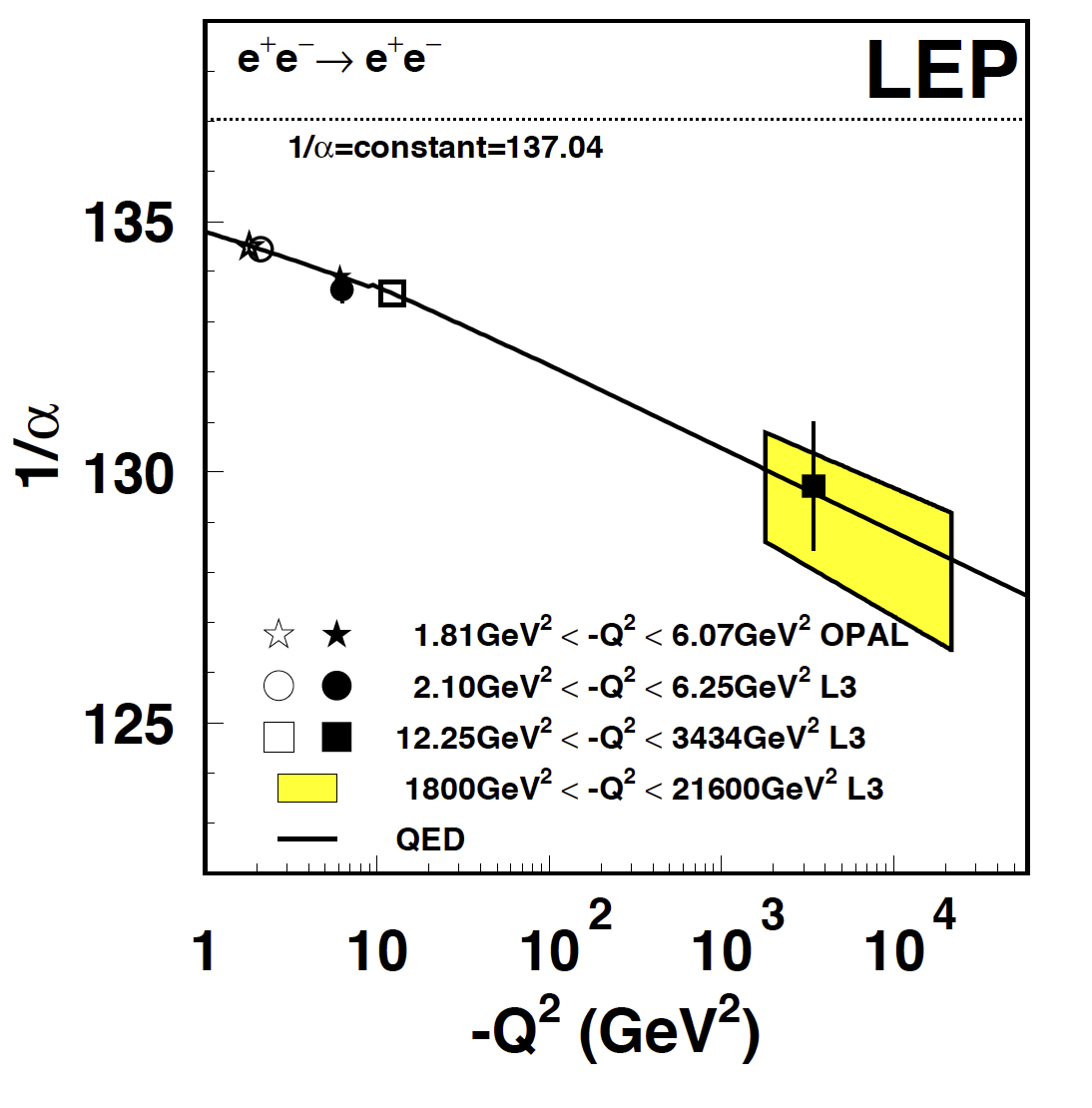}\hspace*{8mm}
   \includegraphics[width=0.47\textwidth,height=0.45\textwidth]{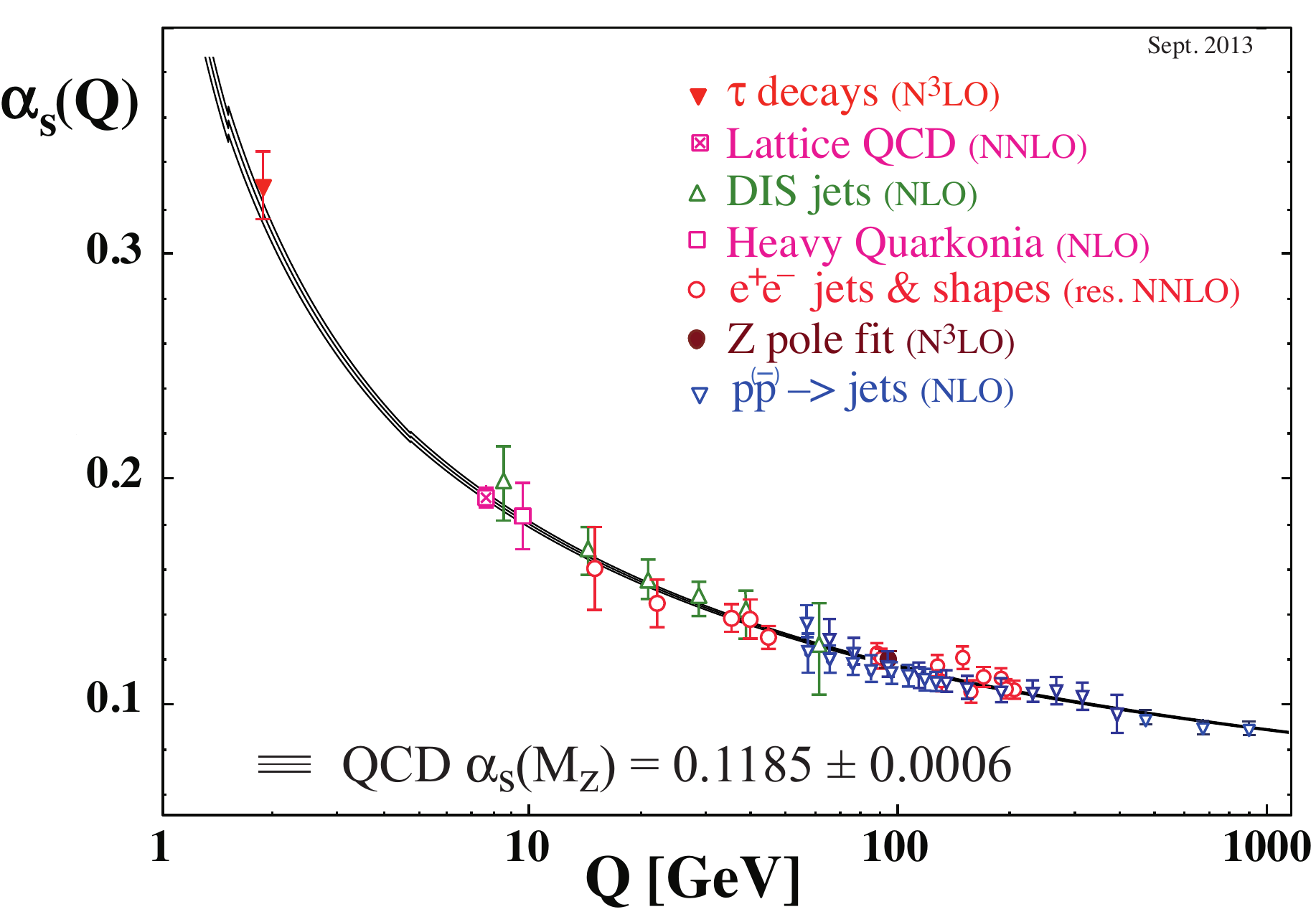}\vspace*{-1mm}
\center{\hspace*{10mm}(a)\hspace*{0.5\textwidth}(b)}
    \caption[]{Running of the electroweak \cite{Mele:2006ji} (a) and strong coupling \cite{Agashe:2014kda} (b) in comparison with the expected running from the RGEs.} \label{f11}
  \end{center}
\end{figure}
\subsection{Running of the b quark mass}\label{s10c}
A bare quark is surrounded by a cloud of gluons, which increases its mass in an energy dependent way. The energy dependence can be calculated by taking into account the running of the coupling constant and the scale, at which the quark mass is probed.
For the b quark mass one expects a change from 4.2 GeV at the b mass to 3 GeV at the Z mass. The b quark mass can be measured by a comparison of the 3-jet rate for b quarks and light quarks\cite{Bernreuther:1997jn,Rodrigo:1997gy,Bilenky:1998nk}, since the b mass effect reduces the cross section by about 5\% \cite{Abreu:1998ih}. 
Comparing the LEP value with the measurements at low energy clearly shows the running \cite{Abreu:1998ih,Barate:2000ab,Abbiendi:2001tw}, see Fig. \ref{f10}b.
 \subsection{Determination of the strong coupling constant}\label{s10d}
Gluon radiation from quarks increases the hadronic \Z\ cross section  by a factor $1+\alpha_s/\pi + ...\approx 1.04$, where the dots indicate the higher order corrections, known up to   $\alpha_s^4$ \cite{Baikov:2012er}.  A precise determination of the cross section allows to extract the strong coupling constant at the \Z\ scale. The  hadronic peak cross section $\sigma^0_{\mbox{\scriptsize{had}}}$ can be  determined either by normalizing to the luminosity or to the leptonic cross section. In the latter case one determines $\Rl$, the ratio of the hadronic and leptonic decay width of the \Z\ boson. The different normalizations yield different  values of the strong coupling constant:  $\alpha_s=0.1154\pm 0.0040$ and $\alpha_s=0.1225\pm 0.0037$, if one uses $\sigma^0_{\mbox{\scriptsize{had}}}$  or  $\Rl$, respectively.   
 Here only $M_Z$, $\Gamma_{\mbox{\scriptsize{tot}}}$ and $\sigma_{\mbox{\scriptsize{had}}}^0$ from all LEP experiments are used in the fit \cite{deBoer:2003xm}. The low value obtained from the cross section normalized to the luminosity is correlated with the low value of the number of neutrino generations, determined as  $N_\nu=2.982(8)$, which is 2.3$\sigma$ below the expected value of 3 neutrino generations. The error is dominated by the common theoretical error on the luminosity, as discussed before.  In contrast, the ratio $\Rl$ does not depend  on the luminosity.
If we require  the number of neutrino generations to  be three, this is most easily obtained by changing the common Bhabha cross section for all LEP experiments by 0.15\% (3$\sigma$), which  leads to $\alpha_s=0.1196\pm 0.0040$, a value close to $\alpha_s=0.1225\pm 0.0037$  from $\Rl$ and also close to the value from the  ratio of the hadronic and leptonic widths of the $\tau$ lepton, $R_\tau$, which yields  $\alpha_s=0.1197 \pm 0.0016$ \cite{Agashe:2014kda}. 
These $\alpha_s$ values are slightly above the world average  of $\alpha_s=0.1185\pm 0.0006$, quoted in the Partice Data Book. However, this value is dominated by the lattice calculations, for which the correlations between the different groups were not taken into account. Instead, only the weighted average was taken, implying that the groups estimating the systematic error from the ``window'' problem conservatively \cite{Aoki:2009tf}, have a small weight. The window problem is, stated simply, the problem of transferring the strong coupling from the non-perturbative regime of fitted quark masses, as used in lattice calculations, to the $\overline{MS}$ scheme, which relies on a perturbative expansion. If one would take the spread in the values from the different lattice calculations as a window for the correct values, as is done in the $\alpha_s$ determination from the $\tau$-data, the error would be a factor three larger, implying consistency between all measurements.

\section{Gauge Coupling Unification}\label{s11}
Shortly after the first high statistics data from LEP became available the gauge couplings were determined with unprecedented precision and by using renorma\-lization group equations (RGEs) \cite{Wilson:1973jj} the couplings can be extrapolated up to high energies. If second order effects are included, one has to consider the interactions between Yukawa and gauge couplings as well as the running of the SUSY- and Higgs masses, which leads to a set of coupled differential equations. They can  be solved  numerically, see Ref. \cite{deBoer:1994dg} for a compilation of the many RGEs and references therein.
However, the second order effects are small and in first order the running of the coupling constants as function of the energy scale $Q$ is proportional to $1/\beta \log(Q^2)$, so the inverse of the coupling constant versus $\log(Q^2)$ is a straight line with a slope given by the $\beta$ coefficient of the RGE.
The fine structure constant is calculated from the RGE to change  from 1/137.035999074 at low energy to 1/(127.940 $\pm$ 0.014) at LEP I energies, which agrees with data, as shown in Fig. \ref{f11}a \cite{Mele:2006ji}.  Also the running of the strong coupling constant agrees with data, as shown in Fig. \ref{f11}b \cite{Agashe:2014kda}.
 One can obtain the gauge couplings at the Z scale from
$\alpha_1=(5/3)g^{\prime 2}/(4\pi)=5\alpha/(3\cos^2\theta_W)$,
$\alpha_2= g^2/(4\pi)=\alpha/\sin^2\theta_W$,
$\alpha_3= g_s^2/(4\pi)$,
where $g'~,g$ and $g_s$ are the $U(1)$, $SU(2)$ and $SU(3)$ coupling constants\footnote{The couplings are usually given in the $\MSbar$ scheme. However, for SUSY  the dimensional reduction $\DRbar$ scheme is  more appropriate \cite{Antoniadis:1982vr}.  
It has the advantage that the three gauge couplings meet exactly at one point.
                     The $\MSbar$ and $\DRbar$ couplings differ by a small offset
$1/{\alpha_i^{\DRbar}}={1/}{\alpha_i^{\MSbar}}-{C_i}/{12\pi}$,
where  $C_i=N$ for SU($N$) and 0 for U(1), so $\alpha_1$ stays the same.}.
\begin{figure}
  \begin{center}
    \includegraphics[width=1.02\textwidth]{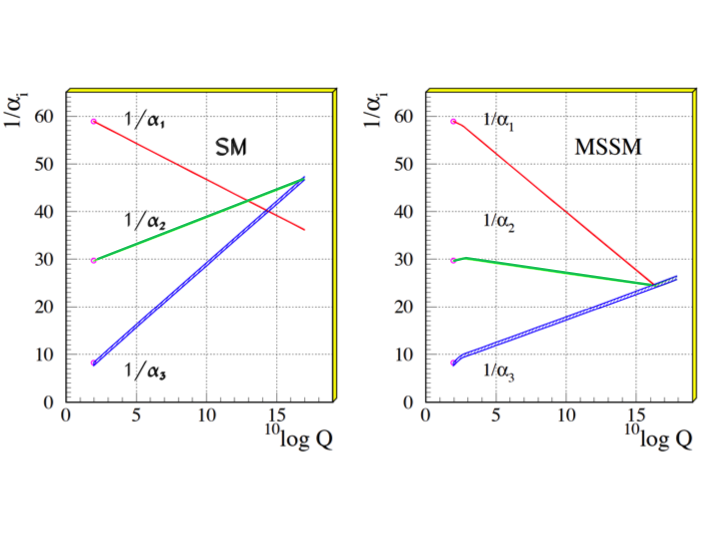}\vspace*{-5mm}
\center{\hspace*{10mm}(a)\hspace*{0.4\textwidth}(b)}
    \caption[]{The running of the couplings in the SM (a) and MSSM (b) using the second order RGEs with a proper threshhold correction for each SUSY particle\cite{deBoer:2003xm}. Note that the running in the MSSM is slower leading to an order of magnitude larger GUT scale, which is  consistent with the limit on the proton lifetime ($\propto 1/M_{GUT}^4$). The widths of the lines correspond to the experimental errors. } \label{f12}
  \end{center}
\end{figure}
The connection between the first two couplings and the electroweak mixing angle can be obtained from  Fig. \ref{f2}b.
   The factor   $5/3$ in the definition of $\alpha_1$ is needed for 
the proper normalization of the gauge groups, whose operators are required to be represented by traceless matrices, see e.g. Ref. \cite{deBoer:1994dg}.
 Fig.  \ref{f12}a demonstrates that the gauge coupling constants  do not meet in a single point,  at least of the RGEs from the SM are used\footnote{The tests for unification in the SM were done before LEP in 1987 by Amaldi et al. \cite{Amaldi:1987fu}, but the precision of the couplings was not high enough to exclude unification in the SM. Amaldi suggested to repeat the analysis with the new LEP data, which showed that within the SM unification is excluded. We found that it is  perfectly possible in Supersymmetry.}. However,  the running of the couplings changes, if  one includes SUSY particles in the loops.  Allowing  the SUSY mass scale and GUT scale to be free parameters in a fit requiring unification allows to derive these scales and their uncertainties \cite{Amaldi:1991cn}. Perfect unification is possible at a scale above $10^{16}$ GeV, which is consistent with the lower limits on the proton lifetime,  as shown in Fig. \ref{f12}b,  in agreement with unification results from other groups  \cite{Ellis:1991wk,Giunti:1991ta,Langacker:1991an,Ellis:1991ri,deBoer:2003xm,Carena:1993ag,Bagger:1995bw}.
Such a unification is by no means trivial, even  from the naive argument, that two lines always meet, so three lines can always brought to a single meeting point with one additional free parameter, like the SUSY mass scale. However, since new mass scales effect all three couplings simultaneously, unification is only reached in rare cases \cite{Amaldi:1991zx}. E.g. a fourth family with an arbitrary mass scale changes all slopes by the same amount, thus never leading to unification. 

\begin{figure}
  \begin{center}
    \includegraphics[width=0.46\textwidth,height=0.26\textheight]{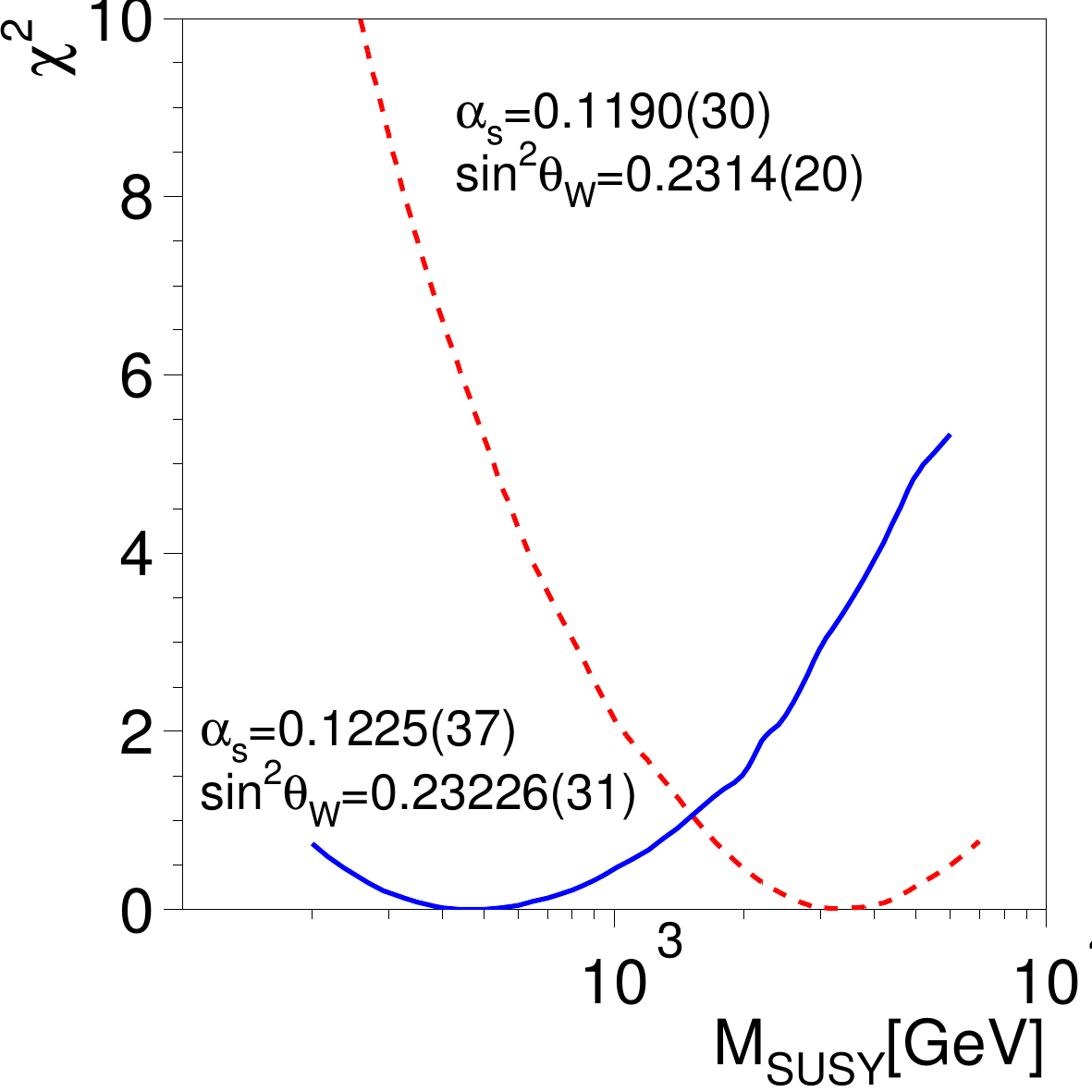}\hspace*{7mm}
    \includegraphics[width=0.46\textwidth,height=0.26\textheight]{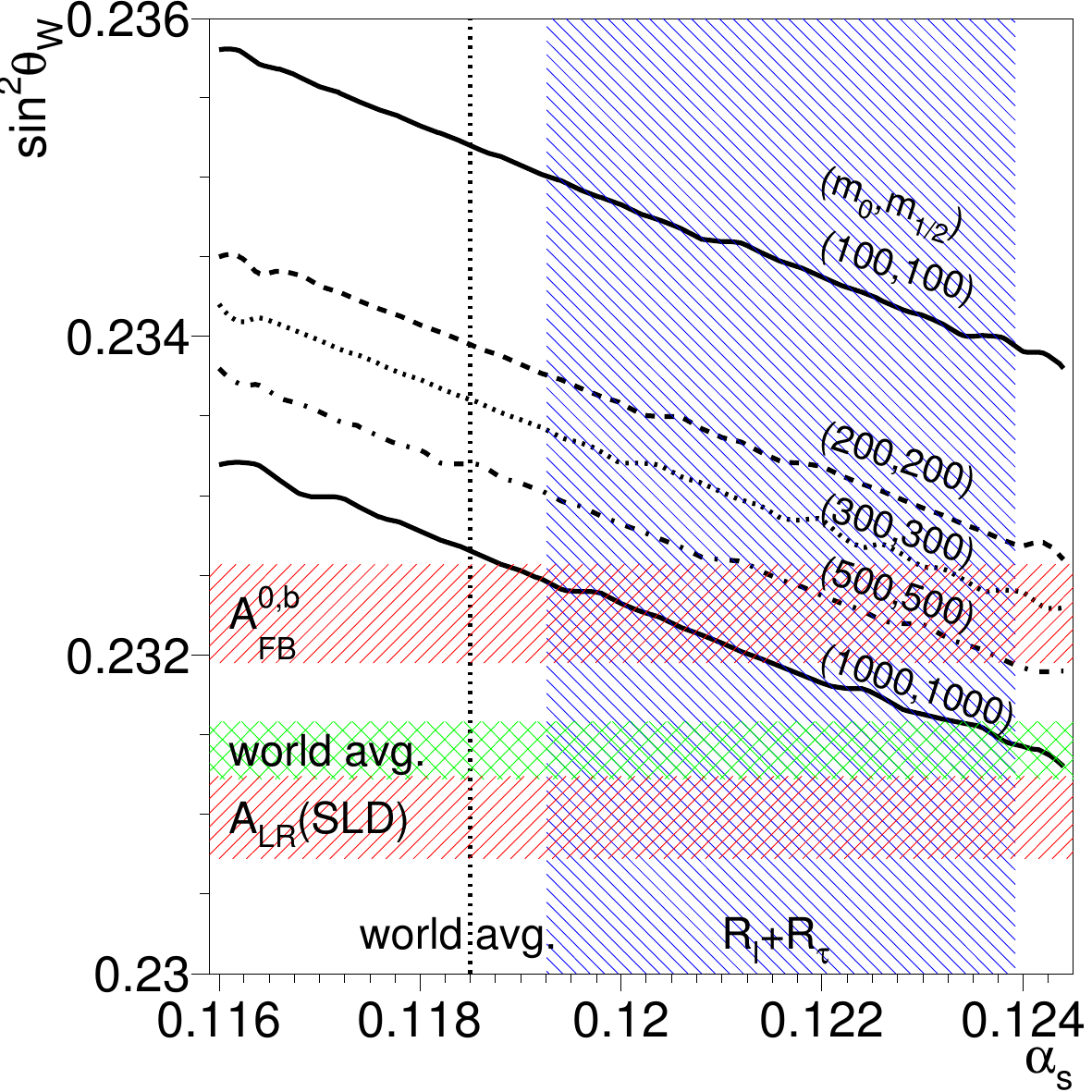}\vspace*{-3mm}
\center{\hspace*{10mm}(a)\hspace*{0.5\textwidth}(b)}
    \caption[]{(a): The $\chi^2$ distribution of $M_{\mbox{\scriptsize{SUSY}}}$ \cite{deBoer:2003xm}. The two different sets of $\alpha_s(M_Z)$ and \sinw\  yield quite different SUSY masses needed for unification, as indicated by the minima. (b): The inclined lines, with the SUSY masses of the CMSSM indicated in brackets in GeV, 
      yield perfect gauge unification.  The horizontal shaded bands indicate the \sinw\ measurements from LEP and SLC, respectively, while the vertically shaded band indicates the value of the strong coupling constant from $\Rl$ and $R_\tau$. These values are above the world average, but well motivated (see text) and they lead more easily to unification. From Ref. \cite{deBoer:2003xm} } \label{f13}
  \end{center}
\end{figure}

The SUSY mass scale depends  on the values of the couplings at the Z scale, as can be seen from the minima of the $\chi^2$ distributions in  Fig. \ref{f13}a for slightly different couplings leading to variations in the SUSY scale from 0.5 to 3.5 TeV.
Hence, the values of $\alpha_s$, $\sinw$ and $M_{SUSY}$ are  correlated. The combination of these three parameters yielding perfect unification are indicated by the diagonal lines for given values of  $M_{SUSY}$ in the  $\alpha_s$, $\sinw$ plane in  Fig. \ref{f13}b \cite{deBoer:2003xm}. Here the full second order RGEs were used with step functions in the beta coefficient at the threshold for each SUSY particle using the particle spectrum from the constrained minimal supersymmetric model (CMSSM), which assumes equal masses $m_0$   ($m_{1/2}$)  for the spin 0  (1/2) particles at the GUT scale. Low energy mass differences originate from the running of the masses from the GUT scale to the low energy scale, taken to be the mass of the SUSY particle. 
  The horizontal bands indicate the value of \sinw~ from $A_{FB}^{0,b}$ and $A_{LR}$.  For \sinw~ from $A_{LR}$ no unification is possible with the central value of $\alpha_s$. However, this of \sinw~ value is inconsistent with the value of \sinw~ from $A_{FB}^{0,b}$ at the $3\sigma$ level (Sect.\ref{s8}). With \sinw~ from $A_{FB}^{0,b}$ unification is possible for $\alpha_s \approx 0.12$ and $M_{SUSY} \gt 1 TeV$. 
These values are consistent with the  $\alpha_s$ value from  observables not depending on the luminosity,  ($\Rl$ and $R_\tau$ indicated by the shaded vertical band in Fig. \ref{f13}b) and present limits on $M_{SUSY}$ from LHC.  Clearly, new data from a future Z-factory would be highly welcome  to settle the minor, but important  discrepancies in $\alpha_s$  and $\sinw$ displayed in Fig. \ref{f13}b.
\section{Summary}\label{s12}
The electroweak precision data from LEP and SLC have provided a remarkable verification of the quantum structure of the SM. Not only the masses of the top quark  and Higgs boson mass could be inferred from the quantum corrections, but also a possible hint for the SM being part  of a Grand Unified Theory was obtained from the running of the gauge couplings in  case the symmetries of the SM are extended by  another symmetry, namely Supersymmetry. 
Supersymmetry solves several shortcomings of the SM (see e.g. Refs \cite{Haber:1984rc,deBoer:1994dg,Martin:1997ns,Kazakov:2010qn} for reviews): i) Electroweak symmetry breaking (EWSB) does not need to be introduced ad hoc, but is induced via radiative corrections;  ii) EWSB predicts a SM-like Higgs boson mass below 130 GeV; iii)    EWSB explains the large  difference between the GUT and electroweak scale, because of the slow running of the Higgs mass terms from positive to negative values; iv) EWSB  requires the top quark mass to be between 140 and 190 GeV for a correct running  of these Higgs mass terms; v) The quadratic divergences in the loop corrections of the SM disappear in Supersymmetry, because of the cancellations between an equal number of fermions and bosons in the loops; vi) The mass ratio of bottom quark over tau lepton is predicted in SUSY, if one presumes Yukawa coupling unification at the GUT scale; vii)  The lightest SUSY particle is a perfect DM candidate, since it is expected to be stable with a self-annihilation cross section  of the right order of magnitude to provide the correct relic density. 

The only troublesome question: where are all the predicted SUSY particles? LHC has excluded squarks and gluinos below the TeV scale. However, as shown in Fig. \ref{f13}a, gauge unification for SUSY masses up to several TeV is perfectly possible. Also the argument that for heavier SUSY masses the cancellation of the quadratic divergences is impacted, is only qualitative. Anyway, the squarks and gluinos are expected to be the heaviest particles because of the gluon clouds surrounding them, so the gauginos and additional Higgs particles may be considerably lighter. These have only weak production cross sections at the LHC, so we do  not have the sensitivity, even if the energy might be sufficient. E.g. for the associated WZ production in the 3-lepton channel the LHC has typically produced 2500 events per experiment for the present luminosity of abour 20/fb at 8 TeV. Assuming the SUSY partners to be a factor four heavier reduces the cross section roughly by a factor $1/M^4$ or more than two orders of magnitude, bringing them to the edge of discovery.   Even at the full LHC energy  and an integrated luminosity of 3000/fb the discovery reach for charginos will  only be 800 GeV \cite{cms3l}. This integrated luminosity can  be reached around 2030, but of course, nothing may be found, either because the SUSY particles are still  heavier or Nature may have found  ways different from Supersymmetry to circumvent the shortcomings of the SM.
\section{Acknowledgements}
 I thank   Ugo Amaldi, Jens Erler, Klaus Hamacher, Hans K\"uhn,  Herwig Schopper, Greg Snow, Dmitri Kazakov and Wilbur Venus  for useful discussions and comments.

,\providecommand{\href}[2]{#2}\begingroup\raggedright\endgroup
\end{document}

\end{document}

%% file: commands.tex
\newlength{\dslashwidth}

\newcommand{\sinw}{\ensuremath{\sin^2\theta_W}}

\newcommand{\MSbar}{\overline{MS}}
\newcommand{\DRbar}{\overline{DR}}

\newcommand{\TeV}{\ensuremath{\mathrm{Te\kern -0.1em V}}}
\newcommand{\GeV}{\ensuremath{\mathrm{Ge\kern -0.1em V}}}
\newcommand{\MeV}{\ensuremath{\mathrm{Me\kern -0.1em V}}}

\newcommand{\mca}[2]{\multicolumn{#1}{|c|}{#2}}

\newcommand{\antibar}[1]{\ensuremath{\mathrm{#1\overline{#1}}}}

\newcommand{\ee}{\ensuremath{\mathrm{e}^+\mathrm{e}^-}}

\newcommand{\qq}{\antibar{q}}
\newcommand{\ff}{\antibar{f}}
\newcommand{\Afbzb}{\ensuremath{A^{0,\,{\mathrm{b}}}_{\mathrm{FB}}}}

\newcommand{\calL}{\ensuremath{\mathcal{L}}}

\newcommand{\ALR}{\ensuremath{A_{\mathrm{LR}}}}

\newcommand{\dalhad}{\Delta\alpha^{(5)}_{\mathrm{had}}(\MZ^2)}

\newcommand{\MW}{\ensuremath{M_{\mathrm{W}}}}
\newcommand{\MZ}{\ensuremath{M_{\mathrm{Z}}}}

\newcommand{\MH}{\ensuremath{M_{\mathrm{H}}}}

\newcommand{\GW}{\ensuremath{\Gamma_{\mathrm{W}}}}
\newcommand{\GZ}{\ensuremath{\Gamma_{\mathrm{Z}}}}

\newcommand{\Gqq}{\ensuremath{\Gamma_{\qq}}}

\newcommand{\Gll}{\Gamma_{\ell\ell}}

\newcommand{\Z}{\ensuremath{\mathrm{Z}}}

\newcommand{\GF}{G_{\mathrm{F}}}

\newcommand{\thw}{\theta_{\mathrm{W}}}
\newcommand{\swsq}{\sin^2\thw}
\newcommand{\cwsq}{\cos^2\thw}

\newcommand{\eeff}{\ensuremath{\ee\rightarrow\ff}}

\newcommand{\cost}{\mbox{$\cos\theta$}}

\newcommand{\Rz}{R^0}

\newcommand{\Rl}{\Rz_{\ell}}

\newcommand{\gt}{\raisebox{0.2ex}{$>$}}

\def\ifmath#1{\relax\ifmmode #1\else $#1$\fi}%
\def\TeV{\ifmmode {\mathrm{ Te\kern -0.1em V}}\else
                   \textrm{Te\kern -0.1em V}\fi}%
\def\GeV{\ifmmode {\mathrm{ Ge\kern -0.1em V}}\else
                   \textrm{Ge\kern -0.1em V}\fi}%
\def\MeV{\ifmmode {\mathrm{ Me\kern -0.1em V}}\else
                   \textrm{Me\kern -0.1em V}\fi}%
\def\keV{\ifmmode {\mathrm{ ke\kern -0.1em V}}\else
                   \textrm{ke\kern -0.1em V}\fi}%
\def\eV{\ifmmode  {\mathrm{ e\kern -0.1em V}}\else
                   \textrm{e\kern -0.1em V}\fi}%

\def\mca#1#2 {\multicolumn{#1}{|c|}{#2}}
\newcommand{\mt}{M_t}

%% file: Submitted_arxiv_20.9.2015.bbl
\begin{thebibliography}{100}%
\makeatletter
\providecommand{\hrefCMSnoop }[0]{\@secondoftwo}%
\makeatother

\bibitem{nobelall}
\hrefCMSnoop {} {} \textit{
  http://www.nobelprize.org/nobel\_prizes/physics/laureates/year/name-lecture.pdf
  (or html),} (1957-2013) {where name is is one of the following: yang, lee,
  schwinger, feynman, tomonaga, gellmann, glashow, salam, weinberg, thooft,
  veltman, gross, politzer, wilczek, kobayashi, maskawa, nambu, englert, higgs,
  ting, richter, fitch, cronin, meer, rubbia, lederman, schwartz, steinberger,
  friedman, kendall, taylor, perl, koshiba, alvarez, davis, charpak}.

\bibitem{Higgs:1964ia}
\hrefCMSnoop {} {P.~W. Higgs, ``{Broken symmetries, massless particles and
  gauge fields}'',} \textit{ Phys.Lett.} \textbf{ 12} (1964)
132--133.

\bibitem{Higgs:1964pj}
\hrefCMSnoop {} {P.~W. Higgs, ``{Broken Symmetries and the Masses of Gauge
  Bosons}'',} \textit{ Phys.Rev.Lett.} \textbf{ 13} (1964)
508--509.

\bibitem{Englert:1964et}
\hrefCMSnoop {} {F.~Englert and R.~Brout, ``{Broken Symmetry and the Mass of
  Gauge Vector Mesons}'',} \textit{ Phys.Rev.Lett.} \textbf{ 13} (1964)
321--323.

\bibitem{Guralnik:1964eu}
\hrefCMSnoop {} {G.~Guralnik, C.~Hagen, and T.~Kibble, ``{Global Conservation
  Laws and Massless Particles}'',} \textit{ Phys.Rev.Lett.} \textbf{ 13} (1964)
585--587.

\bibitem{'tHooft:1972fi}
\hrefCMSnoop {} {G.~'t~Hooft and M.~Veltman, ``{Regularization and
  Renormalization of Gauge Fields}'',} \textit{ Nucl.Phys.} \textbf{ B44}
  (1972)
189--213.

\bibitem{Georgi:1974sy}
\hrefCMSnoop {} {H.~Georgi and S.~Glashow, ``{Unity of All Elementary Particle
  Forces}'',} \textit{ Phys.Rev.Lett.} \textbf{ 32} (1974)
438--441.

\bibitem{Georgi:1980ic}
\hrefCMSnoop {} {H.~Georgi and S.~Glashow, ``{Unified Theory of Elementary
  Particle Forces}'',} \textit{ Phys.Today} \textbf{ 33N9} (1980)
30--39.

\bibitem{Georgi:1974yf}
\hrefCMSnoop {} {H.~Georgi, H.~R. Quinn, and S.~Weinberg, ``{Hierarchy of
  Interactions in Unified Gauge Theories}'',} \textit{ Phys.Rev.Lett.} \textbf{
  33} (1974)
451--454.

\bibitem{McGrew:1999nd}
\hrefCMSnoop {} {C.~McGrew, R.~Becker-Szendy, C.~Bratton{ et~al.}, ``{Search
  for nucleon decay using the IMB-3 detector}'',} \textit{ Phys.Rev.} \textbf{
  D59} (1999)
052004.

\bibitem{Abe:2014mwa}
\hrefCMSnoop {} {{ Super-Kamiokande Collaboration}, ``{Search for
  proton decay via $p\to\nu K^+$ using 260ââkilotonÂ·year data of
  Super-Kamiokande}'',} \textit{ Phys.Rev.} \textbf{ D90} (2014), no.~7,
  072005,
\href{http://www.arXiv.org/abs/1408.1195}{\texttt{ arXiv:1408.1195}}.

\bibitem{Marciano:1981un}
\hrefCMSnoop {} {W.~J. Marciano and G.~Senjanovic, ``{Predictions of
  Supersymmetric Grand Unified Theories}'',} \textit{ Phys.Rev.} \textbf{ D25}
  (1982)
3092.

\bibitem{Amaldi:1991cn}
\hrefCMSnoop {} {U.~Amaldi, W.~de~Boer, and H.~F{\"u}rstenau, ``{Comparison of
  grand unified theories with electroweak and strong coupling constants
  measured at LEP}'',} \textit{ Phys.Lett.} \textbf{ B260} (1991)
447--455.

\bibitem{Ross:1991qv}
\hrefCMSnoop {} {G.~G. Ross, ``{Evidence of supersymmetry}'',} \textit{ Nature}
  \textbf{ 352} (1991)
21--22.

\bibitem{Hamilton:1991qu}
\hrefCMSnoop {} {D.~Hamilton, ``{A Tentative vote for supersymmetry: Do new
  measurements offer indirect support for an elegant attempt to unify
  fundamental forces of nature?}'',} \textit{ Science} \textbf{ 253} (1991)
272.

\bibitem{Dimopoulos:1991au}
\hrefCMSnoop {} {S.~Dimopoulos, S.~Raby, and F.~Wilczek, ``{Unification of
  couplings}'',} \textit{ Phys.Today} \textbf{ 44N10} (1991)
25--33.

\bibitem{Ramond:2014qla}
\hrefCMSnoop {} {P.~Ramond, ``{SUSY: The Early Years (1966-1976)}'',} \textit{
  Eur.Phys.J.} \textbf{ C74} (2014) 2698,
\href{http://www.arXiv.org/abs/1401.5977}{\texttt{ arXiv:1401.5977}}.

\bibitem{Jungman:1995df}
\hrefCMSnoop {} {G.~Jungman, M.~Kamionkowski, and K.~Griest, ``{Supersymmetric
  dark matter}'',} \textit{ Phys.Rept.} \textbf{ 267} (1996) 195--373,
  \href{http://www.arXiv.org/abs/hep-ph/9506380}{\texttt{
  arXiv:hep-ph/9506380}}.

\bibitem{Ade:2013zuv}
\hrefCMSnoop {} {{ Planck Collaboration}, ``{Planck 2013 results.
  XVI. Cosmological parameters}'',} \textit{ Astron.Astrophys.} \textbf{ 571}
  (2014) A16,
\href{http://www.arXiv.org/abs/1303.5076}{\texttt{ arXiv:1303.5076}}.

\bibitem{Haber:1984rc}
\hrefCMSnoop {} {H.~E. Haber and G.~L. Kane, ``{The Search for Supersymmetry:
  Probing Physics Beyond the Standard Model}'',} \textit{ Phys.Rept.} \textbf{
  117} (1985)
75--263.

\bibitem{deBoer:1994dg}
\hrefCMSnoop {} {W.~de~Boer, ``{Grand unified theories and supersymmetry in
  particle physics and cosmology}'',} \textit{ Prog.Part.Nucl.Phys.} \textbf{
  33} (1994) 201--302,
\href{http://www.arXiv.org/abs/hep-ph/9402266}{\texttt{ arXiv:hep-ph/9402266}}.

\bibitem{Martin:1997ns}
\hrefCMSnoop {} {S.~P. Martin, ``{A Supersymmetry primer}'',} \textit{
  Perspectives on supersymmetry II, Ed. G. Kane} (1997)
\href{http://www.arXiv.org/abs/hep-ph/9709356}{\texttt{ arXiv:hep-ph/9709356}}.

\bibitem{Kazakov:2010qn}
\hrefCMSnoop {} {D.~Kazakov, ``{Supersymmetry on the Run: LHC and Dark
  Matter}'',} \textit{ Nucl.Phys. Proc.Suppl.} \textbf{ 203-204} (2010) 118,
  \href{http://www.arXiv.org/abs/1010.5419}{\texttt{ arXiv:1010.5419}}.

\bibitem{Bertone:2010zza}
G.~Bertone, ed., ``{Particle Dark Matter: Observations, Models and Searches}''.
\newblock Cambridge, UK: Univ. Pr.,
2010.
\newblock

\bibitem{Weinberg:1972tu}
\hrefCMSnoop {} {S.~Weinberg, ``{Effects of a neutral intermediate boson in
  semileptonic processes}'',} \textit{ Phys.Rev.} \textbf{ D5} (1972)
1412--1417.

\bibitem{Darriulat:2004nf}
\hrefCMSnoop {} {P.~Darriulat, ``{The discovery of the W and Z, a personal
  recollection}'',} \textit{ Eur.Phys.J.} \textbf{ C34} (2004)
33--40.

\bibitem{Camilleri:1976mb}
\hrefCMSnoop {} {L.~Camilleri, D.~Cundy, P.~Darriulat{ et~al.}, ``{Physics with
  Very High-Energy e+ e- Colliding Beams}'',} \textit{
  CERN-YELLOW-REPORT-76-18}
(1976).

\bibitem{Abrams:1989ez}
\hrefCMSnoop {} {{ MARK-II Collaboration}, ``{First Measurements
  of Hadronic Decays of the $Z$ Boson}'',} \textit{ Phys.Rev.Lett.} \textbf{
  63} (1989)
1558.

\bibitem{assmann}
\hrefCMSnoop {} {R.~Assmann, M.~Lamont, and S.~Meyers, ``{A Brief History of
  the LEP Collider}'',} \textit{ Nucl.Phys.Proc.Suppl.} \textbf{ 109B} (2002)
17.

\bibitem{Blondel:1987wr}
\hrefCMSnoop {} {A.~Blondel, ``{A Scheme to Measure the Polarization Asymmetry
  at the $Z$ Pole in {LEP}}'',} \textit{ Phys.Lett.} \textbf{ B202} (1988)
145.

\bibitem{Barate:2003sz}
\hrefCMSnoop {} {{ LEP Working Group for Higgs boson searches, ALEPH
  , DELPHI , L3 , OPAL Collaboration}
  , ``{Search for the standard model Higgs boson at LEP}'',}
  \textit{ Phys.Lett.} \textbf{ B565} (2003) 61--75,
\href{http://www.arXiv.org/abs/hep-ex/0306033}{\texttt{ arXiv:hep-ex/0306033}}.

\bibitem{Djouadi:2005gj}
\hrefCMSnoop {} {A.~Djouadi, ``{The Anatomy of electro-weak symmetry breaking.
  II. The Higgs bosons in the minimal supersymmetric model}'',} \textit{
  Phys.Rept.} \textbf{ 459} (2008) 1--241,
\href{http://www.arXiv.org/abs/hep-ph/0503173}{\texttt{ arXiv:hep-ph/0503173}}.

\bibitem{Treille:2002iu}
\hrefCMSnoop {} {D.~Treille, ``{LEP/SLC: What did we expect? What did we
  achieve? A very quick historical review}'',} \textit{ Nucl.Phys.Proc.Suppl.}
  \textbf{ 109B} (2002)
1.

\bibitem{Decamp:1990jra}
\hrefCMSnoop {} {{ ALEPH Collaboration}, ``{ALEPH: A detector for
  electron-positron annnihilations at LEP}'',} \textit{ Nucl.Instrum.Meth.}
  \textbf{ A294} (1990)
121--178.

\bibitem{Aarnio:1990vx}
\hrefCMSnoop {} {{ DELPHI Collaboration}, ``{The DELPHI detector
  at LEP}'',} \textit{ Nucl.Instrum.Meth.} \textbf{ A303} (1991)
233--276.

\bibitem{Adriani:1993gk}
\hrefCMSnoop {} {{ L3 Collaboration}, ``{Results from the L3
  experiment at LEP}'',} \textit{ Phys.Rept.} \textbf{ 236} (1993)
1--146.

\bibitem{Ahmet:1990eg}
\hrefCMSnoop {} {{ OPAL Collaboration}, ``{The OPAL detector at
  LEP}'',} \textit{ Nucl.Instrum.Meth.} \textbf{ A305} (1991)
275--319.

\bibitem{Hartmann:2009zza}
\hrefCMSnoop {} {F.~Hartmann, ``{Evolution of Silicon Sensor Technology in
  Particle Physics}'',} \textit{ Springer Tracts Mod.Phys.} \textbf{ 231}
  (2009)
1--204.

\bibitem{Buskulic:1996hx}
\hrefCMSnoop {} {{ ALEPH Collaboration}, ``{Four jet final state
  production in e+ e- collisions at center-of-mass energies of 130-GeV and
  136-GeV}'',} \textit{ Z.Phys.} \textbf{ C71} (1996)
179--198.

\bibitem{Heister:2001kr}
\hrefCMSnoop {} {{ ALEPH Collaboration}, ``{Final results of the
  searches for neutral Higgs bosons in e+ e- collisions at s**(1/2) up to
  209-GeV}'',} \textit{ Phys.Lett.} \textbf{ B526} (2002) 191--205,
\href{http://www.arXiv.org/abs/hep-ex/0201014}{\texttt{ arXiv:hep-ex/0201014}}.

\bibitem{Abreu:1998ih}
\hrefCMSnoop {} {{ DELPHI Collaboration}, ``{Study of the
  four-jet anomaly observed at LEP center-of-mass energies of 130-GeV and
  136-GeV}'',} \textit{ Phys.Lett.} \textbf{ B448} (1999)
311--319.

\bibitem{Wilson:1973jj}
\hrefCMSnoop {} {K.~Wilson and J.~B. Kogut, ``{The Renormalization group and
  the epsilon expansion}'',} \textit{ Phys.Rept.} \textbf{ 12} (1974)
75--200.

\bibitem{Soding:2010zz}
\hrefCMSnoop {} {P.~S\"oding, ``{On the discovery of the gluon}'',} \textit{
  Eur.Phys.J.} \textbf{ H35} (2010)
3--28.

\bibitem{Veltman:1980fk}
\hrefCMSnoop {} {M.~Veltman, ``{Radiative Corrections to Vector Boson
  Masses}'',} \textit{ Phys.Lett.} \textbf{ B91} (1980)
95.

\bibitem{Sirlin:1980nh}
\hrefCMSnoop {} {A.~Sirlin, ``{Radiative Corrections in the SU(2)-L x U(1)
  Theory: A Simple Renormalization Framework}'',} \textit{ Phys.Rev.} \textbf{
  D22} (1980)
971--981.

\bibitem{Kennedy:1988sn}
\hrefCMSnoop {} {D.~Kennedy and B.~Lynn, ``{Electroweak Radiative Corrections
  with an Effective Lagrangian: Four Fermion Processes}'',} \textit{
  Nucl.Phys.} \textbf{ B322} (1989)
1.

\bibitem{Bardin:1989di}
\hrefCMSnoop {} {D.~Y. Bardin, M.~S. Bilenky, G.~Mitselmakher{ et~al.}, ``{A
  Realistic Approach to the Standard Z Peak}'',} \textit{ Z.Phys.} \textbf{
  C44} (1989)
493.

\bibitem{Hollik:1988ii}
\hrefCMSnoop {} {W.~Hollik, ``{Radiative Corrections in the Standard Model and
  their Role for Precision Tests of the Electroweak Theory}'',} \textit{
  Fortsch.Phys.} \textbf{ 38} (1990)
165--260.

\bibitem{Fanchiotti:1992tu}
\hrefCMSnoop {} {S.~Fanchiotti, B.~A. Kniehl, and A.~Sirlin, ``{Incorporation
  of QCD effects in basic corrections of the electroweak theory}'',} \textit{
  Phys.Rev.} \textbf{ D48} (1993) 307--331,
\href{http://www.arXiv.org/abs/hep-ph/9212285}{\texttt{ arXiv:hep-ph/9212285}}.

\bibitem{Agashe:2014kda}
\hrefCMSnoop {} {{ Particle Data Group} , ``{Review of Particle
  Physics}'',} \textit{ Chin.Phys.} \textbf{ C38} (2014)
090001.

\bibitem{ALEPH:2005ab}
\hrefCMSnoop {} {{ ALEPH , DELPHI , L3 ,
  OPAL , SLD , LEP Electroweak Working Group, SLD
  Electroweak Group, SLD Heavy Flavour Group} , ``{Precision
  electroweak measurements on the $Z$ resonance}'',} \textit{ Phys.Rept.}
  \textbf{ 427} (2006) 257--454,
\href{http://www.arXiv.org/abs/hep-ex/0509008}{\texttt{ arXiv:hep-ex/0509008}}.

\bibitem{Bardin:1999ak}
\hrefCMSnoop {} {D.~Y. Bardin and G.~Passarino,
``{The standard model in the making: Precision study of the electroweak
  interactions}'',}.

\bibitem{Schael:2013ita}
\hrefCMSnoop {} {{ ALEPH, DELPHI, L3, OPAL, LEP Electroweak} ,
  ``{Electroweak Measurements in Electron-Positron Collisions at W-Boson-Pair
  Energies at LEP}'',} \textit{ Phys.Rept.} \textbf{ 532} (2013) 119--244,
\href{http://www.arXiv.org/abs/1302.3415}{\texttt{ arXiv:1302.3415}}.

\bibitem{Renton:2002wy}
\hrefCMSnoop {} {P.~B. Renton, ``{Precision electroweak tests of the standard
  model}'',} \textit{ Rept.Prog.Phys.} \textbf{ 65} (2002) 1271--1330,
\href{http://www.arXiv.org/abs/hep-ph/0206231}{\texttt{ arXiv:hep-ph/0206231}}.

\bibitem{Altarelli:2004fq}
\hrefCMSnoop {} {G.~Altarelli and M.~W. Grunewald, ``{Precision electroweak
  tests of the standard model}'',} \textit{ Phys.Rept.} \textbf{ 403-404}
  (2004) 189--201,
\href{http://www.arXiv.org/abs/hep-ph/0404165}{\texttt{ arXiv:hep-ph/0404165}}.

\bibitem{Hollik:2006hd}
\hrefCMSnoop {} {W.~Hollik, ``{Electroweak theory}'',} \textit{
  J.Phys.Conf.Ser.} \textbf{ 53} (2006)
7--43.

\bibitem{Ward:1998ht}
\hrefCMSnoop {} {B.~Ward, S.~Jadach, M.~Melles{ et~al.}, ``{New results on the
  theoretical precision of the LEP / SLC luminosity}'',} \textit{ Phys.Lett.}
  \textbf{ B450} (1999) 262--266,
\href{http://www.arXiv.org/abs/hep-ph/9811245}{\texttt{ arXiv:hep-ph/9811245}}.

\bibitem{D'Agostini:1989cz}
\hrefCMSnoop {} {G.~D'Agostini, W.~de~Boer, and G.~Grindhammer,
  ``{Determination of $\alpha^- s$ and the $\Z^0$ Mass From Measurements of the
  Total Hadronic Cross-section in $e^+ e^-$ Annihilation}'',} \textit{
  Phys.Lett.} \textbf{ B229} (1989)
160.

\bibitem{ATLAS:2014wva}
\hrefCMSnoop {} {ATLAS, CDF, CMS{ et~al.}, ``{ Collaborations, First
  combination of Tevatron and LHC measurements of the topquark mass}'',}
  \textit{ arXiv:1403.4427}
(2014).

\bibitem{Tishchenko:2012ie}
\hrefCMSnoop {} {{ MuLan Collaboration}, ``{Detailed Report of
  the MuLan Measurement of the Positive Muon Lifetime and Determination of the
  Fermi Constant}'',} \textit{ Phys.Rev.} \textbf{ D87} (2013), no.~5, 052003,
\href{http://www.arXiv.org/abs/1211.0960}{\texttt{ arXiv:1211.0960}}.

\bibitem{Montagna:1998kp}
\hrefCMSnoop {} {G.~Montagna, O.~Nicrosini, F.~Piccinini{ et~al.}, ``{TOPAZ0
  4.0: A New version of a computer program for evaluation of deconvoluted and
  realistic observables at LEP-1 and LEP-2}'',} \textit{ Comput.Phys.Commun.}
  \textbf{ 117} (1999) 278--289,
\href{http://www.arXiv.org/abs/hep-ph/9804211}{\texttt{ arXiv:hep-ph/9804211}}.

\bibitem{Arbuzov:2005ma}
\hrefCMSnoop {} {A.~Arbuzov, M.~Awramik, M.~Czakon{ et~al.}, ``{ZFITTER: A
  Semi-analytical program for fermion pair production in e+ e- annihilation,
  from version 6.21 to version 6.42}'',} \textit{ Comput.Phys.Commun.} \textbf{
  174} (2006) 728--758,
\href{http://www.arXiv.org/abs/hep-ph/0507146}{\texttt{ arXiv:hep-ph/0507146}}.

\bibitem{Erler:1999ug}
J.~Erler, ``{GAPP: Global Analysis of Particle Properties}''.
\newblock \url{http://www.fisica.unam.mx/erler/GAPPP.html}.

\bibitem{Jegerlehner:2009ry}
\hrefCMSnoop {} {F.~Jegerlehner and A.~Nyffeler, ``{The Muon g-2}'',} \textit{
  Phys.Rept.} \textbf{ 477} (2009) 1--110,
\href{http://www.arXiv.org/abs/0902.3360}{\texttt{ arXiv:0902.3360}}.

\bibitem{Heinemeyer:2004gx}
\hrefCMSnoop {} {S.~Heinemeyer, W.~Hollik, and G.~Weiglein, ``{Electroweak
  precision observables in the minimal supersymmetric standard model}'',}
  \textit{ Phys.Rept.} \textbf{ 425} (2006) 265--368,
\href{http://www.arXiv.org/abs/hep-ph/0412214}{\texttt{ arXiv:hep-ph/0412214}}.

\bibitem{Beskidt:2012sk}
\hrefCMSnoop {} {C.~Beskidt, W.~de~Boer, D.~Kazakov{ et~al.}, ``{Constraints on
  Supersymmetry from LHC data on SUSY searches and Higgs bosons combined with
  cosmology and direct dark matter searches}'',} \textit{ Eur.Phys.J.} \textbf{
  C72} (2012) 2166,
\href{http://www.arXiv.org/abs/1207.3185}{\texttt{ arXiv:1207.3185}}.

\bibitem{deBoer:2003xm}
\hrefCMSnoop {} {W.~de~Boer and C.~Sander, ``{Global electroweak fits and gauge
  coupling unification}'',} \textit{ Phys.Lett.} \textbf{ B585} (2004)
  276--286,
\href{http://www.arXiv.org/abs/hep-ph/0307049}{\texttt{ arXiv:hep-ph/0307049}}.

\bibitem{Djouadi:2005gi}
\hrefCMSnoop {} {A.~Djouadi, ``{The Anatomy of electro-weak symmetry breaking.
  I: The Higgs boson in the standard model}'',} \textit{ Phys.Rept.} \textbf{
  457} (2008) 1--216,
\href{http://www.arXiv.org/abs/hep-ph/0503172}{\texttt{ arXiv:hep-ph/0503172}}.

\bibitem{Abdallah:2005cy}
\hrefCMSnoop {} {{ DELPHI Collaboration}, ``{Charged particle
  multiplicity in three-jet events and two-gluon systems}'',} \textit{
  Eur.Phys.J.} \textbf{ C44} (2005) 311--331,
\href{http://www.arXiv.org/abs/hep-ex/0510025}{\texttt{ arXiv:hep-ex/0510025}}.

\bibitem{Zerwas:2004ng}
\hrefCMSnoop {} {P.~Zerwas, ``{W \& Z physics at LEP}'',} \textit{ Eur.Phys.J.}
  \textbf{ C34} (2004)
41--49.

\bibitem{Adeva:1990nw}
\hrefCMSnoop {} {{ L3 Collaboration}, ``{A Test of QCD based on
  four jet events from Z0 decays}'',} \textit{ Phys.Lett.} \textbf{ B248}
  (1990)
227--234.

\bibitem{Abreu:1990ce}
\hrefCMSnoop {} {{ DELPHI Collaboration}, ``{Experimental study
  of the triple gluon vertex}'',} \textit{ Phys.Lett.} \textbf{ B255} (1991)
466--476.

\bibitem{Abreu:1993vk}
\hrefCMSnoop {} {{ DELPHI Collaboration}, ``{Measurement of the
  triple gluon vertex from four - jet events at LEP}'',} \textit{ Z.Phys.}
  \textbf{ C59} (1993)
357--368.

\bibitem{Decamp:1992ip}
\hrefCMSnoop {} {{ ALEPH Collaboration}, ``{Evidence for the
  triple gluon vertex from measurements of the QCD color factors in Z decay
  into four jets}'',} \textit{ Phys.Lett.} \textbf{ B284} (1992)
151--162.

\bibitem{Mele:2006ji}
\hrefCMSnoop {} {S.~Mele, ``{Measurements of the running of the electromagnetic
  coupling at LEP}'',} \textit{ XXVI. Phys. in Collision, Rio de Janeiro,
  hep-ex/0610037} (2006)
  \href{http://www.arXiv.org/abs/hep-ex/0610037}{\texttt{
  arXiv:hep-ex/0610037}}.
  \url{http://www.slac.stanford.edu/econf/C060706/pdf/0610037.pdf}.

\bibitem{Bernreuther:1997jn}
\hrefCMSnoop {} {W.~Bernreuther, A.~Brandenburg, and P.~Uwer,
  ``{Next-to-leading order QCD corrections to three jet cross-sections with
  massive quarks}'',} \textit{ Phys.Rev.Lett.} \textbf{ 79} (1997) 189--192,
\href{http://www.arXiv.org/abs/hep-ph/9703305}{\texttt{ arXiv:hep-ph/9703305}}.

\bibitem{Rodrigo:1997gy}
\hrefCMSnoop {} {G.~Rodrigo, A.~Santamaria, and M.~S. Bilenky, ``{Do the quark
  masses run? Extracting m-bar(b) (m(z)) from LEP data}'',} \textit{
  Phys.Rev.Lett.} \textbf{ 79} (1997) 193--196,
\href{http://www.arXiv.org/abs/hep-ph/9703358}{\texttt{ arXiv:hep-ph/9703358}}.

\bibitem{Bilenky:1998nk}
\hrefCMSnoop {} {M.~S. Bilenky, S.~Caberera, J.~Fuster{ et~al.}, ``{m(b)(m(Z))
  from jet production at the Z peak in the Cambridge algorithm}'',} \textit{
  Phys.Rev.} \textbf{ D60} (1999) 114006,
\href{http://www.arXiv.org/abs/hep-ph/9807489}{\texttt{ arXiv:hep-ph/9807489}}.

\bibitem{Barate:2000ab}
\hrefCMSnoop {} {{ ALEPH Collaboration}, ``{A Measurement of the
  b quark mass from hadronic Z decays}'',} \textit{ Eur.Phys.J.} \textbf{ C18}
  (2000) 1--13,
\href{http://www.arXiv.org/abs/hep-ex/0008013}{\texttt{ arXiv:hep-ex/0008013}}.

\bibitem{Abbiendi:2001tw}
\hrefCMSnoop {} {{ OPAL Collaboration}, ``{Determination of the b
  quark mass at the Z mass scale}'',} \textit{ Eur.Phys.J.} \textbf{ C21}
  (2001) 411--422,
\href{http://www.arXiv.org/abs/hep-ex/0105046}{\texttt{ arXiv:hep-ex/0105046}}.

\bibitem{Baikov:2012er}
\hrefCMSnoop {} {P.~Baikov, K.~Chetyrkin, J.~K{\"u}hn{ et~al.}, ``{Complete ${\cal
  O}(\alpha_s^4)$ QCD Corrections to Hadronic $Z$-Decays}'',} \textit{
  Phys.Rev.Lett.} \textbf{ 108} (2012) 222003,
\href{http://www.arXiv.org/abs/1201.5804}{\texttt{ arXiv:1201.5804}}.

\bibitem{Aoki:2009tf}
\hrefCMSnoop {} {{ PACS-CS Collaboration}, ``{Precise
  determination of the strong coupling constant in N(f) = 2+1 lattice QCD with
  the Schrodinger functional scheme}'',} \textit{ JHEP} \textbf{ 0910} (2009)
  053,
\href{http://www.arXiv.org/abs/0906.3906}{\texttt{ arXiv:0906.3906}}.

\bibitem{Antoniadis:1982vr}
\hrefCMSnoop {} {I.~Antoniadis, C.~Kounnas, and K.~Tamvakis, ``{Simple
  Treatment of Threshold Effects}'',} \textit{ Phys.Lett.} \textbf{ B119}
  (1982)
377--380.

\bibitem{Amaldi:1987fu}
\hrefCMSnoop {} {U.~Amaldi, A.~Bohm, L.~Durkin{ et~al.}, ``{A Comprehensive
  Analysis of Data Pertaining to the Weak Neutral Current and the Intermediate
  Vector Boson Masses}'',} \textit{ Phys.Rev.} \textbf{ D36} (1987)
1385.

\bibitem{Ellis:1991wk}
\hrefCMSnoop {} {J.~R. Ellis, S.~Kelley, and D.~V. Nanopoulos, ``{Probing the
  desert using gauge coupling unification}'',} \textit{ Phys.Lett.} \textbf{
  B260} (1991)
131--137.

\bibitem{Giunti:1991ta}
\hrefCMSnoop {} {C.~Giunti, C.~Kim, and U.~Lee, ``{Running coupling constants
  and grand unification models}'',} \textit{ Mod.Phys.Lett.} \textbf{ A6}
  (1991)
1745--1755.

\bibitem{Langacker:1991an}
\hrefCMSnoop {} {P.~Langacker and M.-x. Luo, ``{Implications of precision
  electroweak experiments for $M_t$, $\rho_{0}$, $\sin^2\theta_W$ and grand
  unification}'',} \textit{ Phys.Rev.} \textbf{ D44} (1991)
817--822.

\bibitem{Ellis:1991ri}
\hrefCMSnoop {} {J.~R. Ellis, S.~Kelley, and D.~V. Nanopoulos, ``{A Detailed
  comparison of LEP data with the predictions of the minimal supersymmetric
  SU(5) GUT}'',} \textit{ Nucl.Phys.} \textbf{ B373} (1992)
55--72.

\bibitem{Carena:1993ag}
\hrefCMSnoop {} {M.~S. Carena, S.~Pokorski, and C.~Wagner, ``{On the
  unification of couplings in the minimal supersymmetric Standard Model}'',}
  \textit{ Nucl.Phys.} \textbf{ B406} (1993) 59--89,
\href{http://www.arXiv.org/abs/hep-ph/9303202}{\texttt{ arXiv:hep-ph/9303202}}.

\bibitem{Bagger:1995bw}
\hrefCMSnoop {} {J.~Bagger, K.~T. Matchev, and D.~Pierce, ``{Precision
  corrections to supersymmetric unification}'',} \textit{ Phys.Lett.} \textbf{
  B348} (1995) 443--450,
\href{http://www.arXiv.org/abs/hep-ph/9501277}{\texttt{ arXiv:hep-ph/9501277}}.

\bibitem{Amaldi:1991zx}
\hrefCMSnoop {} {U.~Amaldi, W.~de~Boer, P.~H. Frampton{ et~al.}, ``{Consistency
  checks of grand unified theories}'',} \textit{ Phys.Lett.} \textbf{ B281}
  (1992)
374--383.

\bibitem{cms3l}
\hrefCMSnoop {} {CMS-, ``{SUSY future analyses for Technical
  Proposal}'',} (2014) CMS--PAS--SUS--14--012.

\end{thebibliography}
